\documentclass{aastex63}

\usepackage{xcolor}

\newcommand{\cawav}{\ion{Ca}{2}~8542\,\AA}
\newcommand{\ca}{\ion{Ca}{2}}
\newcommand{\ha}{H$\alpha$}
\newcommand{\hb}{H$\beta$}
\newcommand{\cmcub}{cm$^{-3}$}

\shorttitle{Solar Off-Limb Flare Ribbons}
\shortauthors{Kuridze et al.}

\begin{document}

\title{Spectral Characteristics and Formation Height of Off-Limb Flare Ribbons}

\correspondingauthor{D. Kuridze}
\email{dak21@aber.ac.uk}

\author[0000-0003-2760-2311]{David Kuridze} 
\affiliation{Department of Physics, Aberystwyth University, Ceredigion, SY23 3BZ, UK}
\affiliation{Abastumani Astrophysical Observatory,  Mount Kanobili, Abastumani, Georgia}

\author[0000-0002-7725-6296]{Mihalis Mathioudakis}
\affiliation{Astrophysics Research Centre, School of Mathematics and Physics, Queen's University Belfast, Belfast BT7 1NN, UK}

\author[0000-0002-5778-2600]{Petr Heinzel}
\affiliation{Astronomical Institute, The Czech Academy of Sciences, 25165 Ond\v{r}ejov, Czech Republic}

\author[0000-0002-7444-7046]{J\'{u}lius Koza} 
\affil{Astronomical Institute, Slovak Academy of Sciences, 059 60 Tatransk\'{a} Lomnica, Slovakia}

\author[0000-0002-6547-5838]{Huw Morgan}
\affiliation{Department of Physics, Aberystwyth University, Ceredigion, SY23 3BZ, UK}

\author[0000-0003-4162-7240]{Ramon Oliver}
\affiliation{Departament de F\'{\i}sica, Universitat de les Illes Balears, E-07122 Palma de Mallorca, Spain}
\affiliation{Institute of Applied Computing \& Community Code (IAC3), UIB, Spain}

\author[0000-0001-7458-1176]{Adam F. Kowalski}
\affiliation{Department of Astrophysical and Planetary Sciences, University of Colorado Boulder, 2000 Colorado Avenue, Boulder, CO 80305, USA}

\author[0000-0003-4227-6809]{Joel C. Allred}
\affiliation{NASA Goddard Space Flight Center, Heliophysics Sciences Division, Code 671, 8800 Greenbelt Road, Greenbelt, MD 20771, USA}

\begin{abstract}
Flare ribbons are bright manifestations of flare energy dissipation in
the lower solar atmosphere. For the first time, we report on
high-resolution imaging spectroscopy observations of flare ribbons
situated off-limb in the \hb\ and \cawav\ lines and make a detailed
comparison with radiative hydrodynamic simulations. Observations of
the X8.2-class solar flare SOL2017-09-10T16:06\,UT obtained with the Swedish
Solar Telescope reveal bright horizontal emission layers in \hb\ line
wing images located near the footpoints of the flare
loops. The apparent separation between the ribbon observed in the
\hb\ wing and the nominal photospheric limb is about $300 - 500$\,km.
The \cawav\ line wing images show much fainter ribbon emissions
located right on the edge of the limb, without clear separation from
the limb. RADYN models are used to investigate synthetic spectral line 
profiles for the flaring atmosphere, and good agreement is found with the observations.
The simulations show that, towards the limb, where the line of sight is
substantially oblique with respect to the vertical direction, the
flaring atmosphere model reproduces the high contrast of the off-limb
\hb\ ribbons and their significant elevation above the photosphere. The ribbons
in the \cawav\ line wing images are located deeper in the lower
solar atmosphere with a lower contrast. A comparison of the height
deposition of electron beam energy and the intensity contribution
function shows that the \hb\ line wing intensities  
can be an useful tracer of flare energy deposition in the lower solar
atmosphere.
\end{abstract}



\section{Introduction}

\begin{figure}
\includegraphics[width=\textwidth]{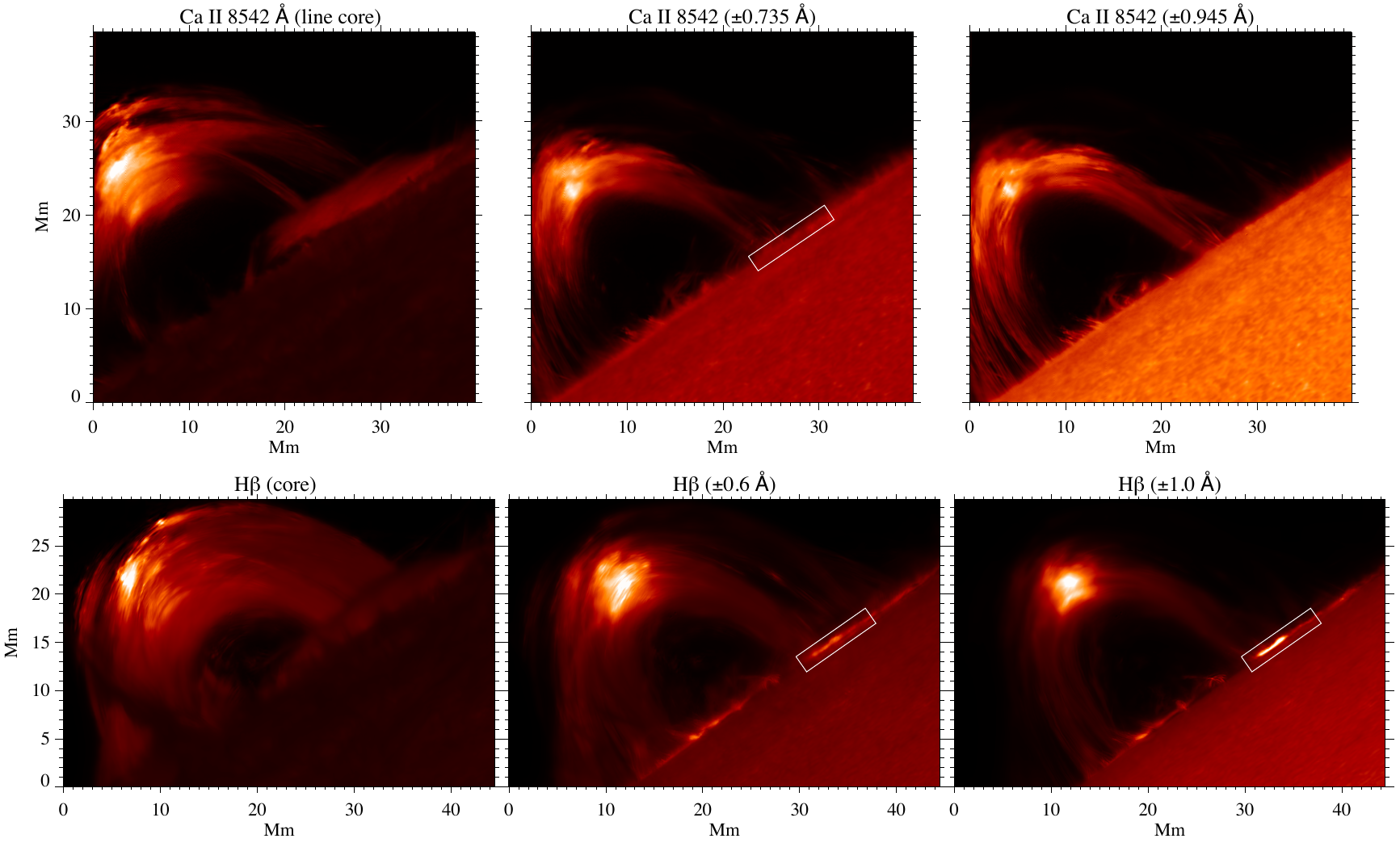}
\caption{An overview of the SST observations of the flare loops on 2017
  September 10 at 16:32\,UT at the western limb. Each image is
  byte-scaled independently. The images shown in Figures~\ref{fig2} and \ref{fig3} are from the same time. Top: CRISP \cawav\ line core
  image (left), composites of near-wing images at $\Delta\lambda = \pm
  0.735$\,\AA\ (middle), and wing images at $\Delta\lambda = \pm
  0.945$\,\AA\ (right). Bottom: CHROMIS \hb\ line core image (left),
  composites of near-wing images at $\Delta\lambda = \pm
  0.6$\,\AA\ (middle), and wing images at $\Delta\lambda = \pm
  1.0$\,\AA\ (right). The white boxes mark the selected region at the flare   
  ribbon that are discussed in the text.}
\label{fig1}
\end{figure}

\begin{figure}
\includegraphics[width=\textwidth]{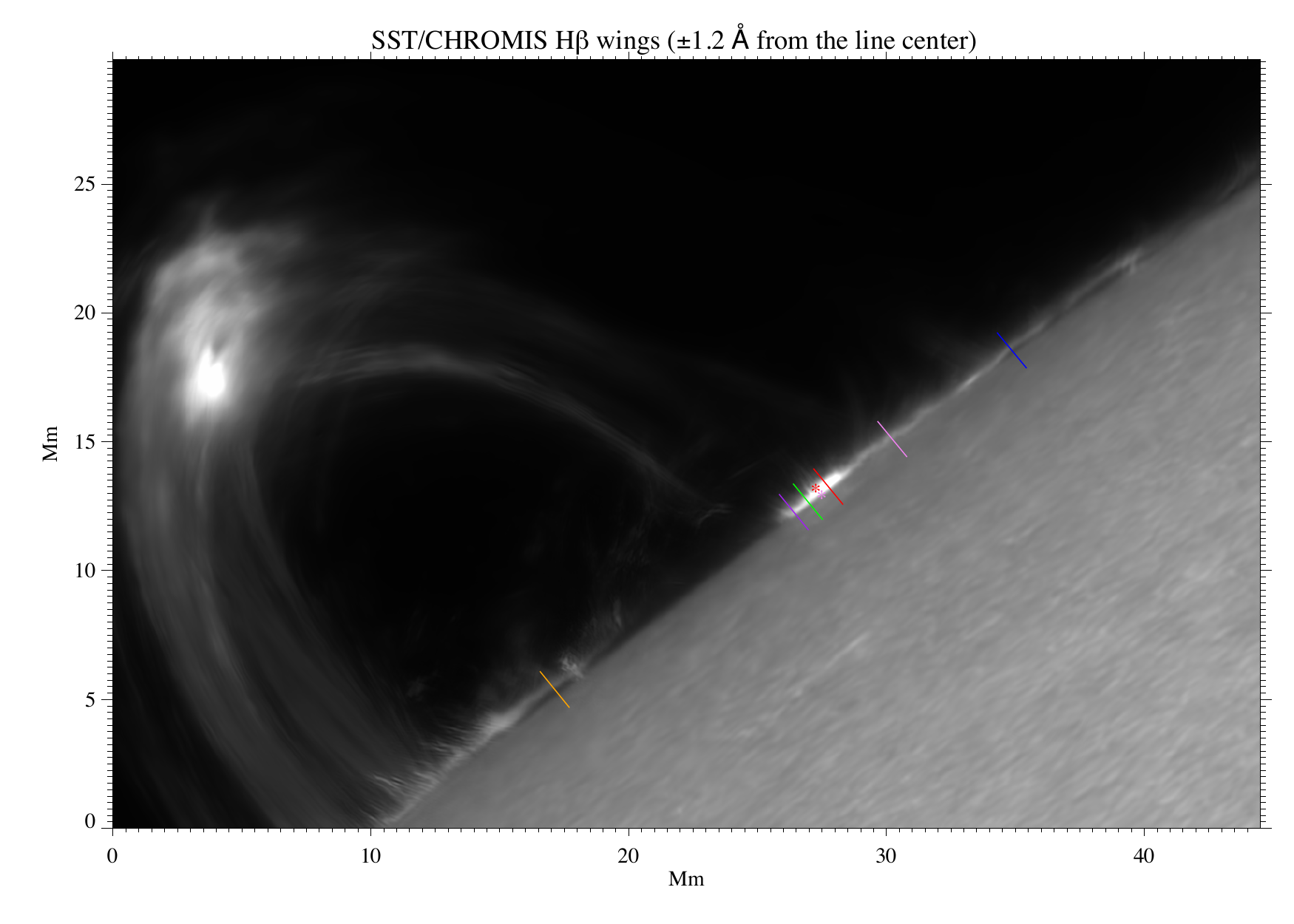}
\includegraphics[width=\textwidth]{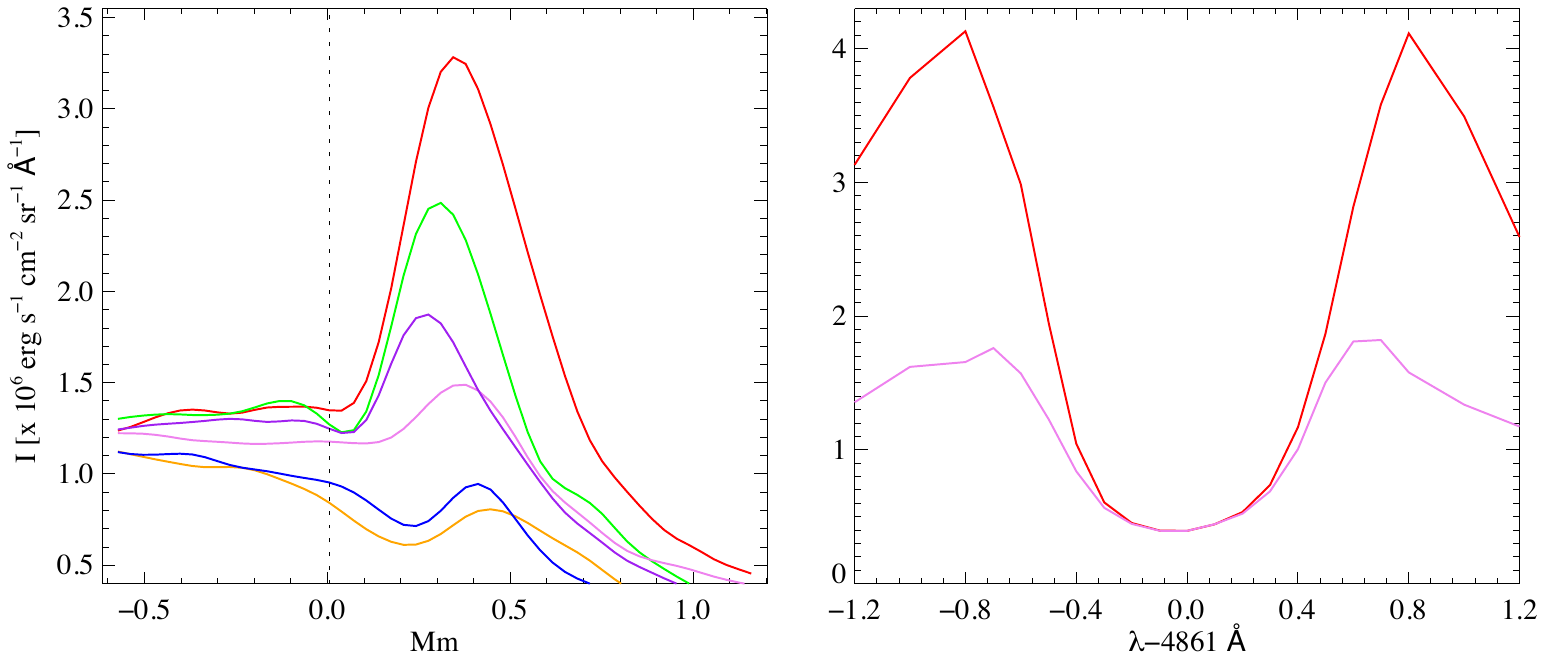}
\caption{Top: Composite of CHROMIS \hb\ far-wing images at
  $\Delta\lambda = \pm 1.2$\,\AA\ of the flare loops and off-limb
  ribbon. Colored dashes correspond to intensity cuts in the
    bottom left panel. Red and magenta asterisks at $(x,y) \sim
    (28,16)$\,Mm mark the locations of the \hb\ line profiles shown
    in the bottom right panel. Bottom left: Intensity cuts $\Delta\lambda=-1.2$\,\AA\ along the
    colored dashes in the top panel. The nominal limb ($h\sim$\,0\,Mm) is
    defined as the middle point of the intensity cut shown by the
    orange line. Bottom right: \hb\ line profiles representing the
    off-limb ribbon emission (red line, red asterisk in the top panel)
    and the dark gap between the ribbon and the limb (magenta line,
    magenta asterisk in the top panel).}
\label{fig2}
\end{figure}

\begin{figure}
\includegraphics[width=\textwidth]{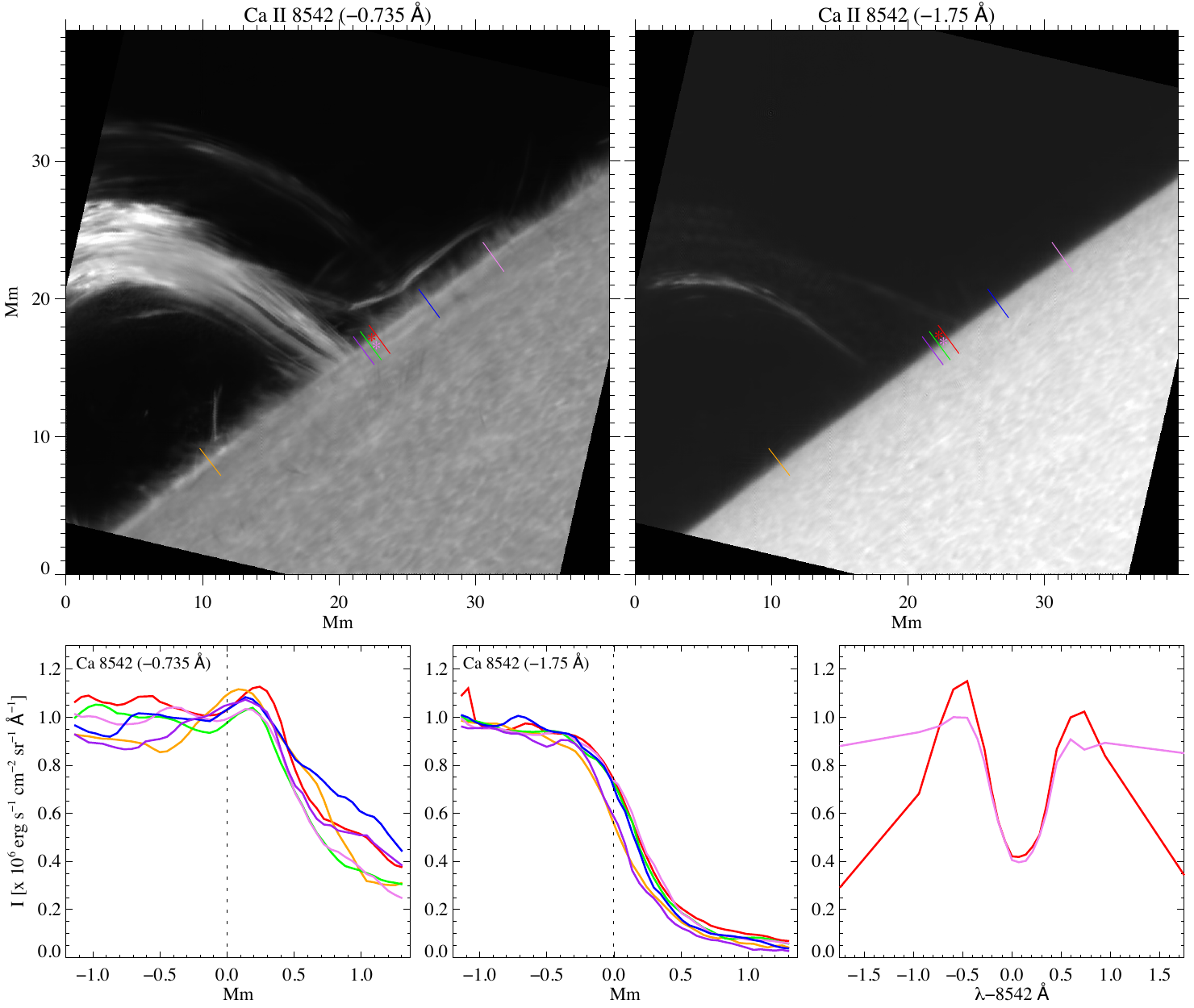}
\caption{Top: CRISP images of the flare loops in the \cawav\ line at
  $\Delta\lambda=-0.735$\,\AA\ (left) and $-1.75$\,\AA\ (right). Red
  and magenta asterisks at $(x,y) \sim (24,16)$\,Mm mark the
  locations of the \cawav\ line profiles shown in the bottom right
  panel. Colored dashes correspond to the intensity cuts shown in the bottom
  left and middle panels. Bottom left and middle: Intensity cuts along
  the colored dashes in the top panels. The nominal limb ($h\sim$\,0\,Mm) is
  defined as an average of all middle points of intensity cuts at
  $\Delta\lambda=-1.75$\,\AA\ (top right panel). Bottom right:
  \cawav\ line profiles representing the ribbon emission (red line,
  red asterisks in the top panels) and the near-ribbon on-disk
  emission (magenta line, magenta asterisk in the top
    panels).}
\label{fig3}
\end{figure}   

\begin{figure}
\includegraphics[width=\textwidth]{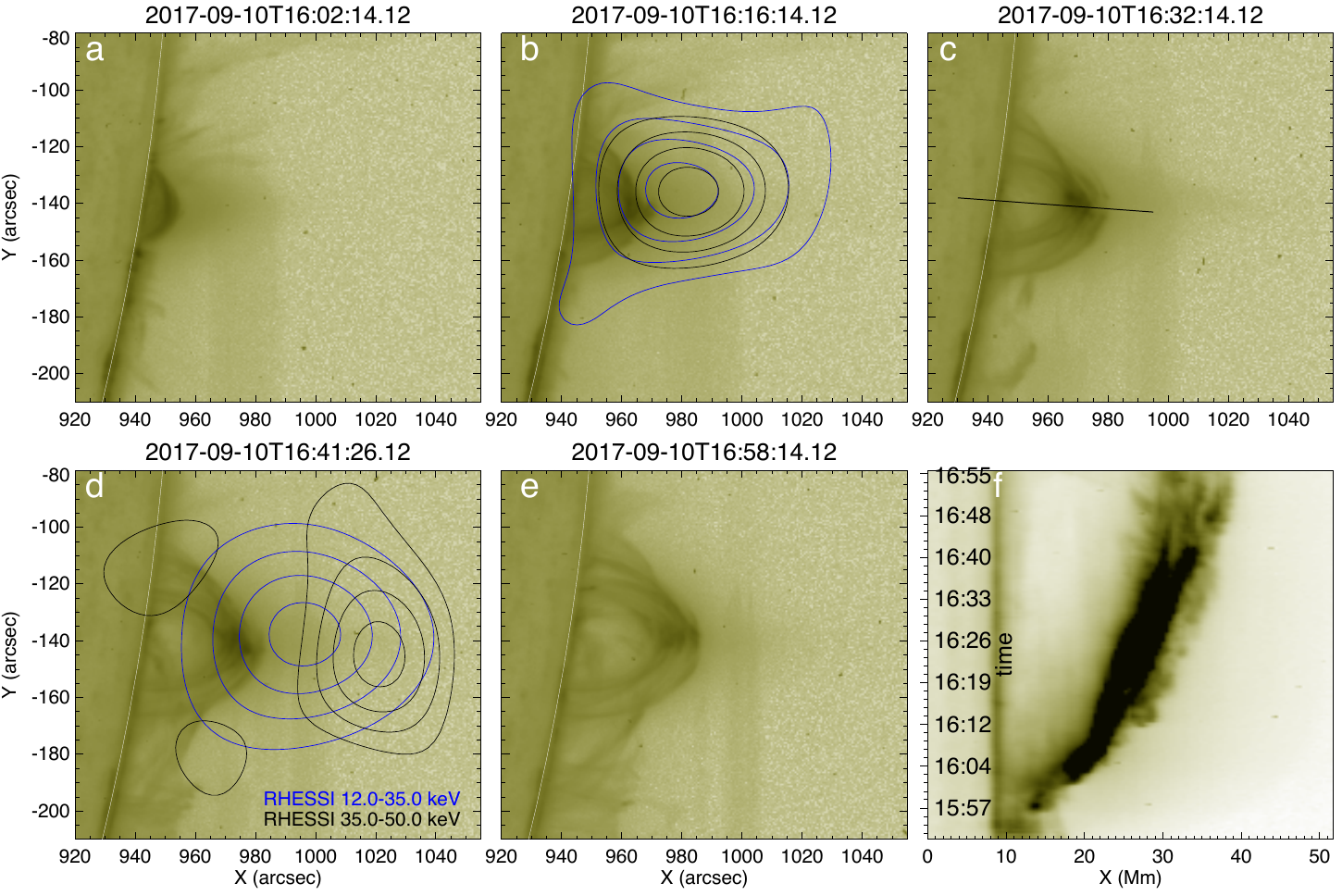}
\caption{{Temporal evolution of the flare loop arcade in 
SDO/AIA 1600 {\AA} of the X8.2 class solar flare loops on 2017 September 10. Images are in log reverse scale so that the brightest emission is in black.
The 16:17 and 16:41 images are overlaid with a RHESSI reconstructed image, shown as blue and black contours at 30\%, 50\%, 70\%, and 90\% 
of the maximum. The radial distance between the surface and the loop arcade top is increasing fast during the first 40 minutes from the main flare peak at 16:06 UT. Panel f is a time-distance diagram of the expanding 
loop arcade as observed in the 1600 {\AA} plotted along the cut presented in panel c. \hb\, ribbon marked with the white box in Figure~\ref{fig1}} is located near [940$''$, $-$165$''$].
The white line marks the position of the limb.}
\label{aia}
\end{figure}

Solar flares are phenomena associated with sudden energy 
releases caused by the rapid reconfiguration of
coronal magnetic fields. A significant amount of the released energy 
is transported along magnetic loops via precipitation of 
accelerated particles and thermal conduction to the lower layers of the 
solar atmosphere \citep{Hirayama1974}. Most of this energy is dissipated
on reaching the dense footpoints of the magnetic loops, giving rise to
flare ribbons -- striking intensity enhancements marking the footpoints
of flare loop arcades \citep{Fletcheretal2011}. Flare ribbons are
observed mainly through the optical, near-infrared, UV, EUV and XEUV
spectral domains. They provide clues to our understanding of the
physics of solar flares and vital diagnostics for the processes of energy dissipation
\citep{Druettetal2017}. Despite tremendous observational and numerical 
modeling efforts over the past few decades, many
aspects of flare ribbon physics, such as their formation height,
spectral characteristics, and radiation mechanisms at different
spectral domains, remain the subject of ongoing debate.

One of the main spectral characteristics of chromospheric ribbons
is the reversed \ha\ line profile with broad, enhanced, and asymmetric
wing emission \citep{IchimotoKurokawa1984,Berlicki2007,Kuridzeetal2015}.  
Other chromospheric spectral lines, such as \ion{Mg}{2}, and \ion{Na}{1}~D1 and D2,
reveal similar flare ribbon characteristics \citep{Kerretal2016,Kuridzeetal2016}. 
The \ion{Ca}{2}~H\,\&\,K and \cawav\ line profiles turn into emission in the 
flaring chromosphere, with or without a central reversal
\citep{AbbettHawley1999,Allredetal2015,Kuridzeetal2015,Kuridzeetal2017,Kuridzeetal2018}.

\citet{Machadoetal1980}
developed semi-empirical models of bright and faint flares that reproduce lines and
continua of the neutral \ion{H}{1}, \ion{Si}{1}, \ion{C}{1}, and singly-ionised  
\ion{Ca}{2} and \ion{Mg}{2}. The models show a substantial
temperature enhancement from the photosphere up to the transition
region. Various static and dynamic models of the flaring chromosphere 
can reproduce the basic spectral characteristics of the observed line profiles
\citep{Canfieldetal1984,Fangetal1993}.  In the majority of flare
simulations the flare energy is transported from reconnection sites
toward the lower atmosphere via electron beams, which are considered
as the most effective energy-transport and heating
mechanism \citep{Brown1971,Hudson1972,Druettetal2017}.
Beam heating can exist for extended periods of time in a flare atmosphere, 
including the gradual flare phase \citep{Tandberg-HanssenEmslie1988}.
\cite{ZharkovaKobylinskii1993} analytically calculated non-thermal 
ionisation and excitation rates of hydrogen in a hydrodynamic atmosphere 
heated by electron beams, demonstrating that the beam heating 
significantly increased the ionisation degree of hydrogen in the lower atmosphere. 
\citet{KasparovaHeinzel2002} also investigated the influence of non-thermal
collisional rates, related to an electron beam, on hydrogen Balmer
line profiles. They show that the intensity of line wings are
significantly enhanced for typical values of the beam energy flux.
The \ion{Ca}{2}~H\,\&\,K and \cawav\ line emissions are also
enhanced, however the line wings are less sensitive to the beam heating 
\citep{Fangetal1993}. 

Dynamic flare models obtained by radiation-hydrodynamic (RHD) 
simulations such as RADYN \citep{CarlssonStein1997, AbbettHawley1999, Allredetal2015}
produce a strong increase of wing emission in the Balmer lines as a
response to electron beam heating \citep{daCosta2008,Kuridzeetal2015}. 
These results suggest that Balmer line wing emission can be a very effective
tracer of flare energy deposition in the lower atmosphere \citep{Canfieldgayley1987}. 
Observations and modeling of flares on dMe stars by \citet{HawleyPettersen1991} 
and \citet{Kowalskietal2010,Kowalskietal2013,Kowalskietal2017} 
reveal that the hydrogen Balmer lines can be extremely broad. More recently,
\citet{DruettZharkova2018} found that beam heating leads to an
increase of Lyman, Balmer, and Paschen continua as well as wing
emission in the Balmer and Paschen lines.
\citet{DruettZharkova2018} also explained the formation of large redshifts 
observed in Balmer lines \citep[see e. g.,][]{IchimotoKurokawa1984}, 
as well as white-light Paschen emission 
generated in solar flares heated by non-thermal electrons. 
\cite{2019A&A...623A..20D} investigated the role of radiative transfer in 
Lyman lines and continuum in maintaining ionisation and white-light emission.
They found that the plasma heating caused by beam electrons led to an increase 
in Lyman line and continuum radiation.

Numerical and semi-empirical models of flares by
\citet{Mauas1990,Mauas2007} show that flare-related perturbations can
affect the formation heights of line cores and wings as well as the
continuum emission originating in the photosphere. Observations of
flare footpoints and ribbons at the solar limb can provide direct
information on the formation height of emission sources in
different spectral lines, which in turn can be crucial to determine the
stopping depth of energetic electrons, where they lose their energy completely.
In contrast to disk flares and their associated ribbons, there is 
a lack of high-resolution observations of off-limb
flare ribbons. \citet{Kruckeretal2015} investigated the formation 
heights of off-limb flare footpoints observed at 6173\,\AA\ in continuum 
images of the {\it Solar Dynamics Observatory's} \citep[SDO;][]{Pesnelletal2012}
{\it Helioseismic and Magnetic Imager} \citep[HMI;][]{Schouetal2012,Scherreretal2012} 
and hard X-ray observations from the {\it Reuven Ramaty High-Energy 
Solar Spectroscopic Imager} \citep[RHESSI;][]{Linetal2002}. 
They show that the 6173\,\AA\ continuum and hard X-ray emissions are
closely correlated in space, time, and intensity, and are formed at
heights of $300 - 450$\,km above the visible solar
limb. \citet{Heinzeletal2017} investigated radiation mechanisms
responsible for enhanced continuum emission at 6173\,\AA\ using HMI
observations of off-limb flares and RHD simulations Flarix \citep{Heinzel2016} of flare heating
by electron beams.  They show that Paschen recombination continuum
dominates in the continuum emission. \cite{DruettZharkova2018} also 
modelled the Paschen continuum emission and investigated its formation depth  
as a function of HXR emission. They found that the height 
distribution of contribution functions for the Paschen continuum  
correlates closely with observations of WL and HXR emission 
reported for limb flares \citep{Martinez2012}.

On 2017 September 10, the Active Region (AR) NOAA 12673 at the western
limb of the Sun produced the SOL2017-09-10T16:06 X8.2-class flare
(Figure~\ref{fig1}), ranked as the second-largest flare of solar
cycle 24, causing significant space weather and heliospheric
effects. The event was observed with the {\it Swedish Solar Telescope}
\citep[SST;][]{Scharmeretal2003a,Scharmeretal2003b} using
high-resolution imaging spectroscopy in the hydrogen
\hb\,4861\,\AA\ Balmer line and in the \cawav\ line. The SST data yielded some unique 
spectropolarimetry which allowed the construction of a
map of the magnetic field of the off-limb flare loops 
\citep{Kuridzeetal2019}. Density diagnostics of the bright apex
of the flare loop arcade are given by \citet{Kozaetal2019}. This paper
presents an analysis of high-resolution imaging spectroscopy data of
off-limb ribbons of the X8.2-class flare acquired by SST. To our
knowledge, high-resolution observations of off-limb flare ribbons in
the hydrogen Balmer lines have not been previously reported. The RHD code
RADYN is used to investigate the formation of the \hb\ and \cawav\ line profiles
and their intensity contribution functions in the flaring atmosphere
involving ribbons at the limb.

\section{Observations and Data Reduction}

AR NOAA 12673 was observed between 16:07:21\,UT and 17:58:37\,UT
on 2017 September 10 when it was close to the western limb, with
the heliocentric coordinates of the center of the SST field of view (FoV) at
(947\arcsec, $-138$\arcsec) initially. Observations were made with the CRisp Imaging
SpectroPolarimeter \citep[CRISP;][]{Scharmer2006,Scharmeretal2008} and
the CHROMospheric Imaging Spectrometer (CHROMIS) instruments, both
based on dual Fabry-P\'{e}rot interferometers (FPI) mounted on SST.
The imaging setup includes a dichroic beamsplitter with the
transmission/reflection edge at 5000\,\AA. CRISP is mounted in the
reflected red beam and CHROMIS in the transmitted blue beam
\citep[Figure~2]{Lofdahletal2018}.

CRISP data is comprised of narrow-band imaging spectropolarimetry in
the \cawav\ line profile sampled from $-1.75$\,\AA\ to
$+1.75$\,\AA\ in 21 line positions $\pm 1.75$, $\pm 0.945$, $\pm
0.735$, $\pm 0.595$, $\pm 0.455$, $\pm 0.35$, $\pm 0.28$, $\pm 0.21$,
$\pm 0.14$, $\pm 0.07$, 0.0\,\AA\ from line center (hereafter, unless
specified otherwise, when referring to the \ca\ line we mean the
\cawav\ line). The \ca\ Stokes {\it I} profiles taken
at 16:32\,UT are used for the analysis; a detailed analysis 
of the Stokes {\it Q, U, V} profiles is presented in
\citet{Kuridzeetal2019}. The CRISP data is processed by the CRISPRED
reduction pipeline \citep{delaCruzRodriguezetal2015} and reconstructed
with Multi-Object Multi-Frame Blind Deconvolution
\citep[MOMFBD;][]{Lofdahl2002,vanNoortetal2005}.

Simultaneous observations were taken with the CHROMIS imaging
spectrometer: a dual FPI observing in the range of $3900-4900$\,\AA. 
The CHROMIS observations comprise narrow-band and wide-band 
spectral imaging in several spectral lines, continua, and 
pseudo-continua \citep[Figure~3, Table~1]{Lofdahletal2018}. 
The CHROMIS \hb\ data taken at 16:32\,UT are used for the analysis. 
The \hb\ line scan consists of 21 profile samples
ranging from $-1.2$\,\AA\ to $+1.2$\,\AA\ at positions $\pm 1.2$, $\pm
1.0$, $\pm 0.8$, $\pm 0.7$, $\pm 0.6$, $\pm 0.5$, $\pm 0.4$, $\pm
0.3$, $\pm 0.2$, $\pm 0.1$, and 0.0\,\AA\ from line center. CHROMIS
data are processed using the CHROMISRED reduction pipeline, which
includes MOMFBD image restoration \citep{Lofdahletal2018}.

The spatial sampling was 0.057$''$pixel$^{-1}$ and 0.0375$''$pixel$^{-1}$  
for the \ca\ and \hb\ lines, respectively. The spatial resolutions were close to 
the diffraction limit of the telescope at this wavelengths, i.e., $\sim$0.215$''$ 
(155 km) and $\sim$0.122$''$ (88 km) for the selected \ca\ and \hb\ images 
in the time series. More details on the observations, data, and
radiometric calibration can be found in \citet{Kuridzeetal2019} and
\citet{Kozaetal2019}. Due to the highly variable and less-than-optimum
seeing, only one CRISP and CHROMIS scan taken at the moment of the
best viewing conditions at $\sim$\,16:32\,UT, i.e. $\sim$\,25\,min after
the flare peak at $\sim$\,16:06\,UT is used in our analysis.

The flare was also observed with the Fermi Gamma-ray Burst 
Monitor \citep[GBM;][]{Meegan2009}. The flare was observed only 
partially with RHESSI \citep{Lin2002} as observations were suspended 
between $\sim$16:17 and 16:40 UT during the satellite's passage through 
the South Atlantic Anomaly. In Section~\ref{HXR} we describe the 
analysis of the Fermi\,GBM and RHESSI datasets. The hard X-ray 
spectral analysis was performed using the Object Spectral Analysis 
Executive \citep[OSPEX;][]{Schwartz2002} to estimate the power 
$P_\mathrm{nth}$ deposited in the chromosphere by the non-thermal
electrons, assuming a thick-target model.

The event was also observed with NASA's Solar Dynamics Observatory 
(SDO) Atmospheric Imaging Assembly \citep[AIA;][]{Lemen2012} 
1600/1700 {\AA} and several Extreme-UV channels as well as the HMI 
\citep{Schouetal2012,Scherreretal2012} instrument in white-light continuum.

\begin{figure}
\includegraphics[width=0.478\textwidth]{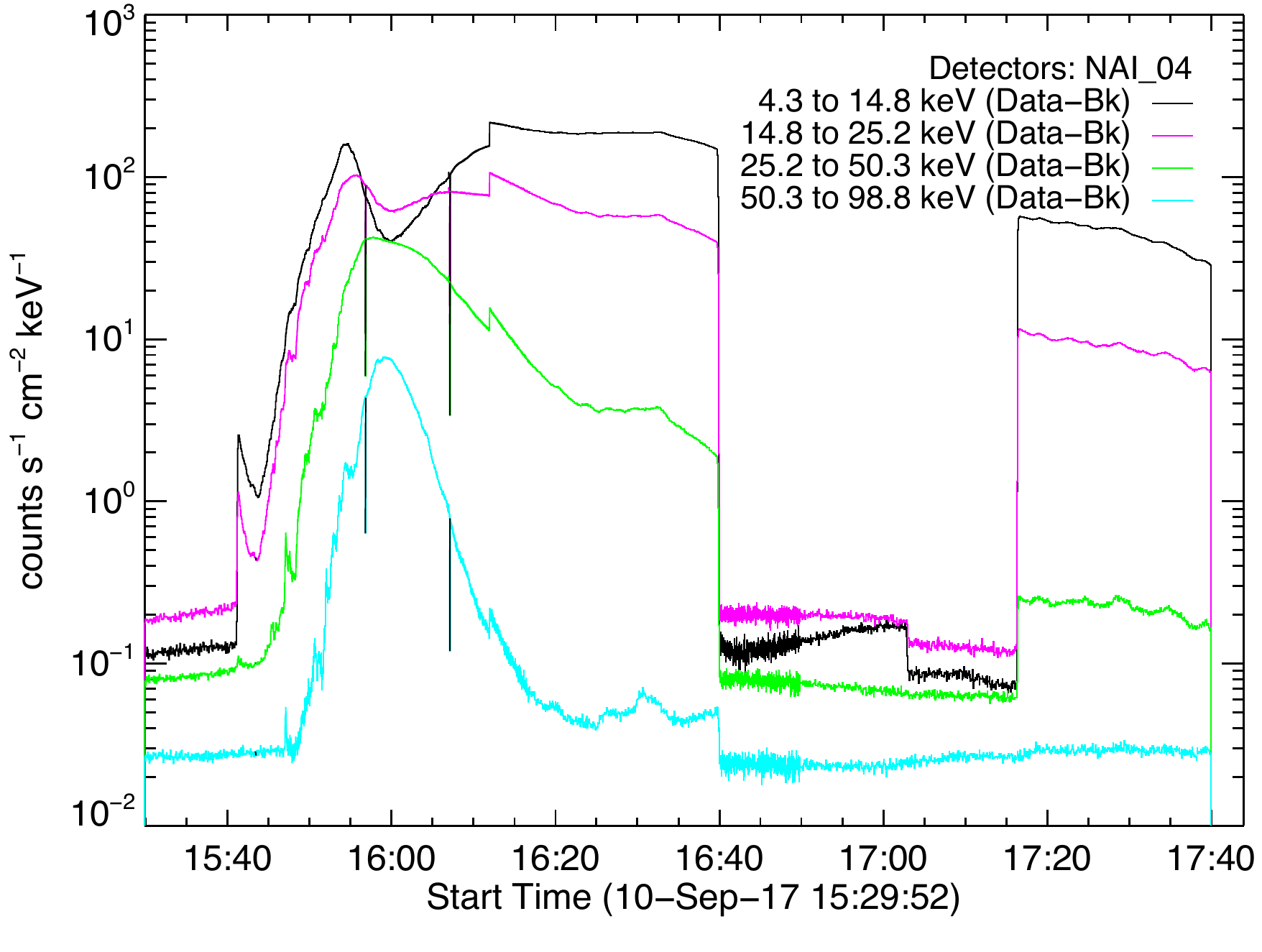}
\includegraphics[width=0.515\textwidth]{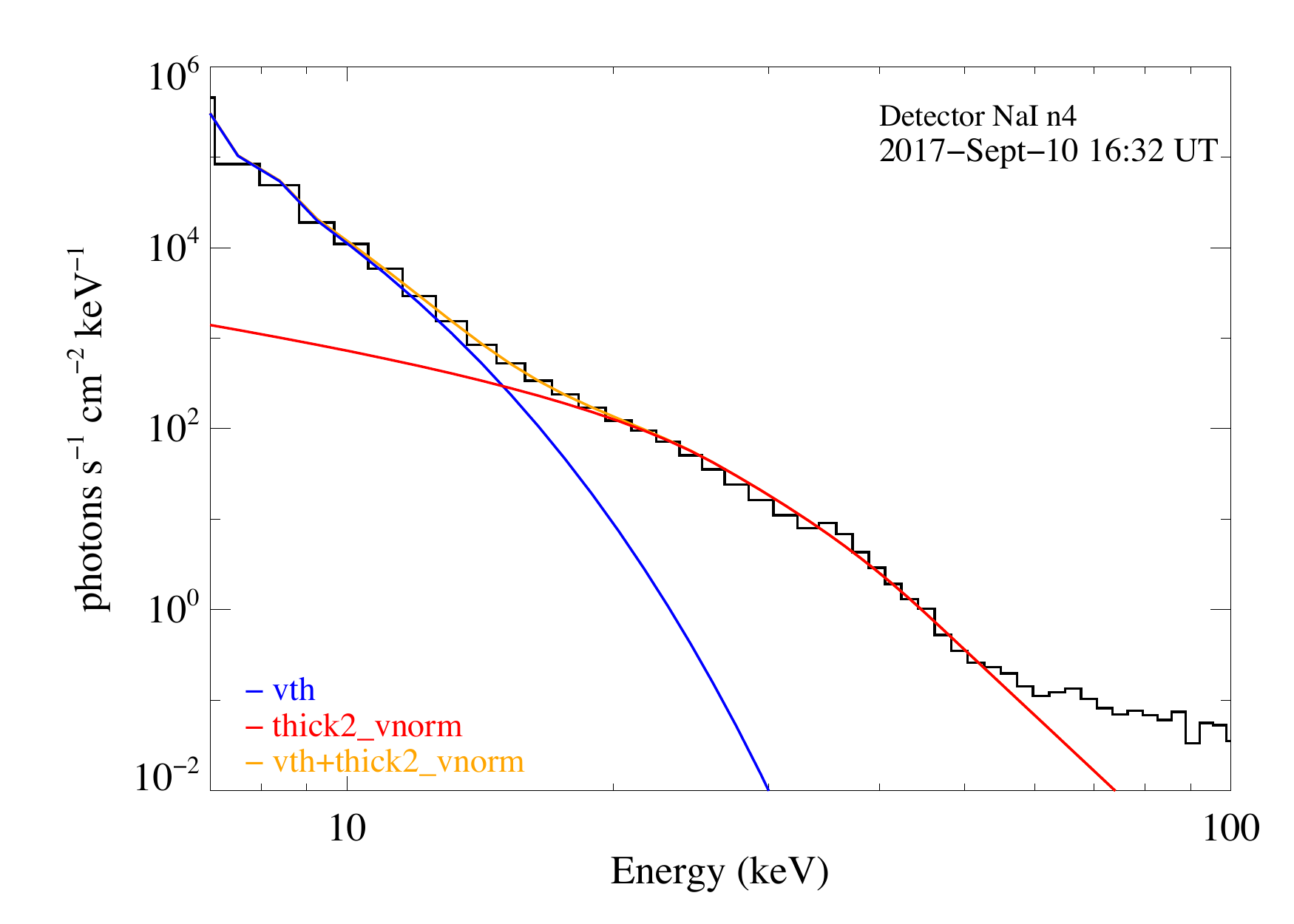}
\caption{Left: Fermi\,GBM background-subtracted X-ray fluxes (Data-Bk) of the X8.2-class flare of
    2017 September 10 from the NaI\,4 detector in the energy bands
    from 4.3\,keV (black) to 98.8\,keV (cyan). Right: X-ray
    background-subtracted photon spectra (Data-Bk, black line) at
    16:32\,UT. The blue, red and orange lines represent the
    fitted thermal \texttt{vth}, non-thermal power-law
    \texttt{thick2\_vnorm}, and total fitting components
    \texttt{vth+thick2\_vnorm}, respectively.}
\label{fermi}
\end{figure}

\section{Analysis and Results}

\subsection{Off-Limb Ribbon Appearance}

Figure~\ref{fig1} shows flare images in the \ca\ and \hb\ line core
and wing positions taken at 16:32\,UT. At that time, the line core and
wing images show well-developed flare loops. Bright narrow rims
at the loop footpoints in both \hb\ wing images closely follow
the limb (Figures~\ref{fig1}: bottom middle and right panels, Figure~\ref{fig2}). 
The right leg of the observed flare arcade in H$\beta$ and \ca\ lines is anchored 
in this brightening (Figure~\ref{fig1}, \ref{fig2}, \& \ref{fig3}). 
The morphology of the flare event suggests that the electron beam propagates 
down along the flare loops from the apex, heating the chromosphere near the 
loop footpoints, manifested as the off limb 
brigtenings seen in H$\beta$ and \ca\ images (white boxes in Figure~\ref{fig1}). 
Therefore, we interpreted them as the off-limb flare ribbons projected on the 
limb and seen from the side. 
SDO/HMI continuum data 
at 6173 {\AA} also show the off-limb enhanced emission as a slipping
ribbon feature \citep{jejcic2018}.
The off-limb brightening located in the ranges 
$\mathrm{x =10-22~Mm~and~y=0 - 7~Mm}$ in Figure~\ref{fig2}
can not be confidently identified as the footpoints of a flare loop arcade. 
Therefore, our focus is on the flare ribbon near the right footpoint of 
the arcade marked with the white box in Figure \ref{fig1} and covering 
the brightest flare kernel. The ribbons are absent in both \hb\ and
\ca\ core images (Figure~\ref{fig1}: left panels).

Figure~\ref{fig2} shows that the ribbons are located above the edge of
the disk, which is separated from the ribbons at some sections by a
dark gap. To investigate the separation between the off-limb ribbons
and the limb, intensities are extracted along short dashes of length 1.8\,Mm\ 
which cross the limb, shown as coloured lines in the top panel of Figure~\ref{fig2}. 
The position of these dashes are chosen to cross the limb, the dark gap, 
and the off-limb ribbons. The intensity profiles are shown in the bottom left
panel of Figure~\ref{fig2}. The position of the 
nominal \hb\ limb, or the nominal $h \sim$\,0\,Mm level, 
is defined as the middle point of the steep intensity gradient 
at $\Delta\lambda =-1.2$\,\AA\ along the orange dash. 
This level varies at different positions
along the limb indicating the presence of inhomogeneous footpoint
heating and also a possible Wilson depression. 
It is known that the limb is uplifted by about 
$\sim$350 km above the base of the photosphere at 
$\tau_{\rm 5000} = 1$ (measured at $\mu=1$) \citep[Table~1]{Lites1983}.
We note that $\mu$ is the cosine of the angle $\theta$ between the 
LoS and the surface normal (heliocentric angle) 
so that $\theta$=0$^\circ$ at disk center and 90$^\circ$ at the limb.  
This off-set is due to an increase of optical depth at 5000 {\AA}
when looking from disk center ($\mu=1$) to limb ($\mu=0$). 
The nominal limb defined above is also expected to be formed at
$\sim$350 km above the $\tau_{\rm 5000} = 1$ at $\mu=1$. 
The apparent separation projected on the plane of sky, 
defined as the distance between maximum intensity of 
the ribbon emission and the nominal limb, 
varies between $300-500$\,km (Figure~\ref{fig2}: bottom
left panel). The estimated vertical extension of the ribbon at \hb\,$\pm1.2$\,\AA,
defined as the average of the full width at half maxima at several
intensity cuts across the ribbon, is around 300\,km.

The bottom right panel of Figure~\ref{fig2} shows the ribbon and gap
\hb\ line profiles from the pixel positions at $(x,y) \sim
(28.5,13)$\,Mm identified in the top panel by the asterisks. 
We assume that these line profiles are observed perpendicular 
($\mu\approx0$) to the propagation 
of the electron beam along the loops anchored at the footpoints.
Both the ribbon and the dark gap \hb\ profiles are reversed. The wing
intensities of the ribbon at $|\Delta\lambda|>0.4$\,\AA\
are strongly enhanced, whereas intensities around line core at
$|\Delta\lambda|<0.4$\,\AA\ coincide with the line core
intensities of the dark gap.

The top panel of Figure~\ref{fig3} shows the \ca\ wing images at 
wavelength positions $\Delta\lambda=-0.735$\,\AA\ and
$-1.75$\,\AA\ from the line center. The intensities along the
colored dashes crossing the limb are shown in the bottom left and
middle panels. The far wing image and corresponding intensity cuts at
$\Delta\lambda=-1.75$\,\AA\ (Figure~\ref{fig3}: top right and bottom
middle panels) do not show any obvious trace of the ribbon. 
Therefore these images are used to define the nominal \ca\ limb 
setting nominal $h \sim$\,0\,Mm, 
as the average of the middle points of the intensity cuts
along the colored dashes crossing the limb. However, the ribbon is
present as a small intensity enhancement at $h \sim$\,0.1 $-$ 0.25\,Mm in the
$\Delta\lambda=-0.735$\,\AA\ wing image and the corresponding intensity
cuts (Figure~\ref{fig3}: top left and bottom left panels). This suggests
that the \ca\ ribbon emission is formed right at the edge of the disk
without any clear separation between the ribbon and the nominal
\ca\ limb. The bottom right panel of Figure~\ref{fig3} shows the ribbon
and near-limb \ca\ line profiles from the pixel positions at $(x,y)
\sim (24,17)$\,Mm identified in the top panels by the asterisks. The
ribbon \ca\ line profile is very broad, with core intensities coinciding 
with the line core intensities of the near-limb profile. 
The enhanced wing intensities within the wavelength ranges ($-1.0, -0.5$)\,\AA\ 
and ($0.5, 1.0$)\,\AA\ with respect to the corresponding wing intensities of the
near-limb profile give rise to a brighter limb rim which is interpreted as the
\ca\ ribbon (Figure~\ref{fig3}: bottom left panel).

\subsection{AIA\,1600 {\AA}, RHESSI and Fermi\,GBM hard X-ray spectra}
\label{HXR}

The imaging spectroscopy observations in the H$\beta$ and \ca\ lines 
were recorded at $\sim$\,16:32\,UT,  
$\sim$\,25\,min after the main flare peak. 
To investigate the strength of non-thermal heating at this time 
we study the morphological evolution of the 
flare arcade and analyse the available Hard X-ray (HXR) data.

Figure~\ref{aia} shows the temporal evolution of the flare loop arcade 
until approximately 1 hour after the flare onset in a cool AIA 1600 {\AA} 
channel. This is the same flare loop arcade covered by the SST FoV, 
however SST and AIA images presented in Figure~\ref{fig1},~\ref{fig2},~\ref{fig3} and 
Figure~\ref{aia}, respectively, do not have the same orientation.
The morphology and the evolution of the flare system indicates that the 
flare energy propagates along the legs of the flare arcade, from the top towards  
the footpoints. As a result the new flare loops are 
formed intensively during the first $\sim$50$-$60 minutes after the main peak. 
This is manifested as an apparent expansion of the flare loop arcade (Figure~\ref{aia}). 
This apparent expansion is a well-known phenomenon 
related to continuous/progressive reconnection, 
where the reconnection X-point re-forms at increasing height. As a result, 
the height of newly reconnected closed loops increases in time \citep{hori1997}.  
An associated effect is the apparent increasing separation of flare ribbons
at the loop footpoints.
Two-dimensional models of solar flare loops from one-dimensional 
hydrodynamic calculations  by \cite{hori1997} demonstrate that the total 
energy released, and the energy release rate, are determined by the 
reconnection rate and hence an apparent expansion of flare loops.  
Images of the flare arcade and time-distance diagram in Figure~\ref{aia} 
clearly show this apparent expansion during about an hour after flare onset, 
suggesting that the non-thermal electron beam heating is still present 
at 16:32 UT, 25 minutes after the main peak.

{HXR emission is a typical signature for the presence of accelerated 
electrons in flares. The Fermi\,GBM sunward-facing NaI\,4 detector 
measured $4.3 -98.8$\,keV X-ray and HXR flux curves are shown in the left 
panel of Figure~\ref{fermi}. The emissions between $\sim 4-50$\,keV 
are significantly increased at $\sim$\,16:32\,UT. The HXR spectrum 
is fitted with a thermal plus thick-target power law model using the OSPEX
function {\tt \verb$thick2_vnorm$} yielding electron beam
parameters such as total power contained in the electron distribution
$P_\mathrm{nth}$, spectral index $\delta$, and the low-energy cutoff
$E_{\rm c}$ of the flare. The right panel of Figure~\ref{fermi} shows an example of 
the NaI\,4 detector spectrum taken at 16:32\,UT.
The fitted thermal and non-thermal components are overplotted 
as the blue and red lines respectively. The fitting to NaI\,1, 2, and 3 
detector data yield parameters which are consistent with those from NaI\,4.
The $\mathrm{\sim20 - 40}$ keV energy range is excluded from the fit 
due to known calibration issues \citep{Kowalskietal2019}.
The spectral results give estimates of $\delta\sim 5.2$, $E_{\rm
c} \sim$26.5\,keV, and $P_\mathrm{nth} \sim7.8\times10^{28}$\,erg\,s$^{-1}$ at
$\sim$\,16:32\,UT (Figure~\ref{fermi}: right panel).

\begin{figure}
\includegraphics[width=\textwidth]{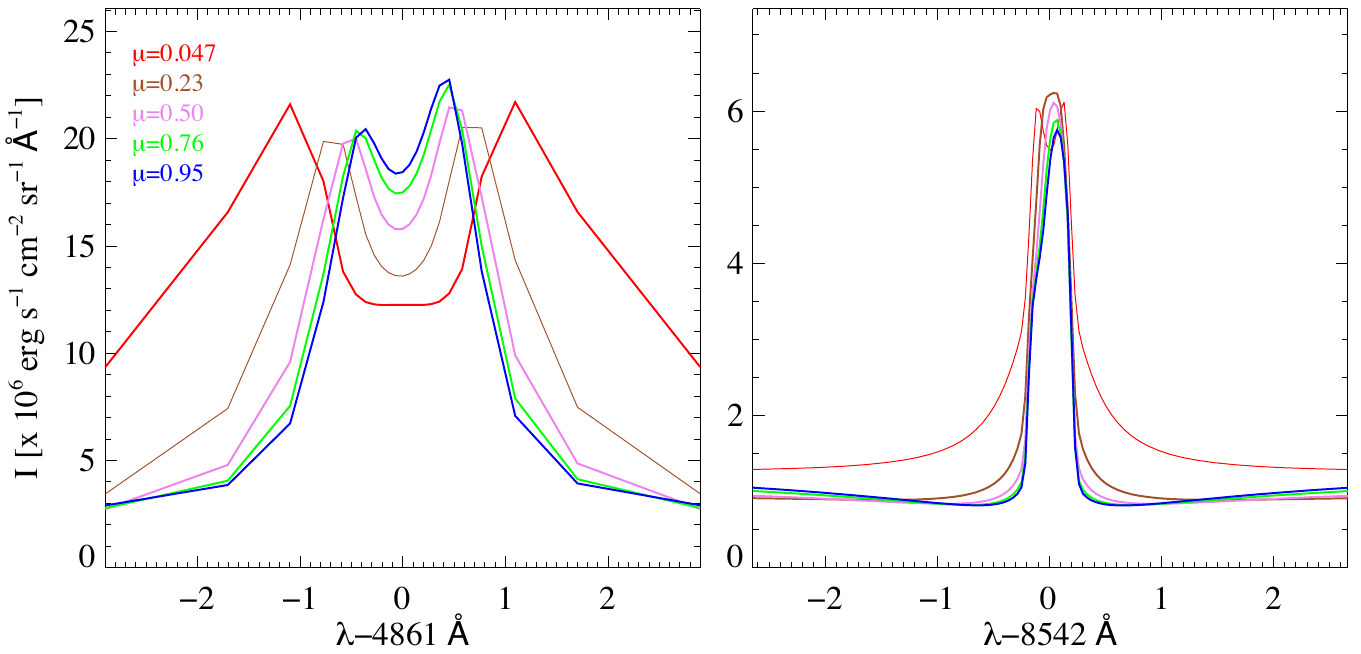}
\caption{Synthetic \hb\ (left) and \cawav\ (right) line profiles from
the RADYN model F11 for five directions representing viewing 
angles from disk centre to the limb.}
\label{fig5}
\end{figure}

Unfortunately, there are no RHESSI imaging spectroscopy data for 
this event between 16:16 - 16:41 UT. Therefore, we can not identify the 
HXR sources at the time of the analysed SST observations and make an accurate 
estimate of the size of the footpoints and the energy flux deposited into the 
chromosphere. Neither can the area of the ribbon be estimated by the far 
wing \hb\ data as the full 2D extension of the ribbons is not seen, 
and they are partially covered by the limb. The reconstructed images for 
the HXR emission \citep[using CLEAN;][detectors 3-8, beam width 
factor of 1]{Hurford2002}, at the closest time to the analysed SST observations, 
are shown as black and blue contours in Figure~\ref{aia}b and \ref{aia}d. 
They indicate the presence of HXR sources at the flare arcade, 
above the arcade, and near the footpoints (Figure~\ref{aia}b and \ref{aia}d).

\subsection{RADYN Code and RHD Model}

To interpret the observational characteristics of the flare ribbon
emissions in the \hb\ and \ca\ lines we use a model of a flaring
atmosphere and synthetic line profiles computed with the RHD code
RADYN \citep{CarlssonStein1997, AbbettHawley1999,Allredetal2015}. 
For a given model of a flare atmosphere, RADYN solves simultaneously
the equations of hydrodynamics, statistical equilibrium, and radiative
transfer under the non-local thermodynamic equilibrium (non-LTE) conditions 
(i.e. departures from LTE) on a 1D adaptive grid to synthesize spectral line
profiles. The flare-related heating is included via a non-thermal
electron beam. RADYN uses five bound levels plus continuum for 
hydrogen, eight levels plus continuum for neutral helium,
five bound levels plus continuum for singly-ionized calcium, and
three bound levels plus continuum for singly-ionized magnesium
with complete frequency redistribution (CRD). RADYN 
uses beam excitation and ionization rates from \cite{Fangetal1993}.
Details on RADYN calculations of non-thermal collisional rates for 
different atomic species are provided in section 4 of \cite{Allredetal2015}.

Several improvements have been made to the RADYN flare code since 
\cite{Allredetal2015}, which are worth noting (they will be 
described further in Allred et al. 2020, in preparation). 
The hydrogen line broadening from \cite{Kowalskietal2017} and 
\cite{Tremblaybergeron2009} has been included in 
the dynamic simulations. The relevant line broadening 
work has been largely completed by \cite{Kowalskietal2017}, where the absorption coefficients in the framework 
of the linear Stark broadening theory of \cite{Vidal1971} are used, as employed 
and modified in \cite{Tremblaybergeron2009} (profiles in 
\cite{Kowalskietal2017} were taken directly from \cite{Tremblaybergeron2009}).  
In \cite{Kowalskietal2017}, the line broadening was incorporated into the 
radiative transfer code RH \citep{Uitenbroek2001}
and in the present model the Doppler-convolved profiles 
from \cite{Tremblaybergeron2009} have been interpolated at each 
temperature and electron density and substituted directly 
into RADYN (except using a 6-level hydrogen atom in RADYN vs. a 20 
level hydrogen atom in RH).  The motivation to replace the \cite{Sutton1978} 
plus Voigt profile method (previously used in RADYN and RH) 
was discussed in \cite{Allredetal2015} and \cite{Kowalski2015}.   
Further details on Stark theory and absorption coefficients 
are given by \cite{Vidal1971}, \cite{Vidal1973}, 
\cite{Tremblaybergeron2009}, and \cite{Kowalskietal2017}.  
The inclusion of hydrogen line Stark profiles in RADYN, including results and comparisons 
for different electron density regimes, will be presented in an 
upcoming work (Kowalski et al. 2020 in prep). 

The QS.SL.HT model is relaxed with the new hydrogen 
broadening, and we choose to use the X-ray backheating formulation 
from \citet{Allredetal2005}, described further in 
Kowalski et al. 2020 (in preparation). The resulting pre-flare 
apex temperature is 1.8 MK, with electron density 
$5 \times 10^9$\,cm$^{-3}$. Finally, a new version of the F-P 
solver is used (Allred et al. 2020, in preparation), which gives a 
moderately smoother electron beam energy deposition profile over 
height in the upper chromosphere. These changes have been 
implemented for the models presented in Graham et al. (2020, under revision). 
The differences in the broadening of the H$\beta$ line profiles at $\mu$=0.047 
between the new and old RADYN models are significant. The new model 
is adopted in this work.

The simulations were performed for a strong electron beam flux with $F
= 10^{11}$\,erg\,cm$^{-2}$\,s$^{-1}$ (an F11 flare), an
isotropic pitch angle distribution in the forward hemisphere
\citep{Allredetal2015}. {A constant heating flux was applied for 20\,s
into a plage-like initial model, closest to QS.SL.HT atmosphere 
described in \cite{Allredetal2015}.} A power-law index and a low 
energy cut-off of $\delta = 4$ and $E_{\rm c} = 25$\,keV, respectively, 
were adopted in the simulation. As mentioned in section~\ref{HXR}, 
the morphology of the analysed flare event cannot be used 
to constrain the electron beam energy flux. Apart from F11, we also investigated the F9 and F10 RADYN models 
(with 100 and 10 times weaker beam energy flux). However, 
these weaker models cannot reproduce the observed enhancement 
of the \ca\ line wing emission. This motivated our choice for the F11 beam 
heating model. Other beam parameters are very close to those 
estimated from the HXR data (section~\ref{HXR}).

The standard continuous reconnection scenario observed 
during the SOL2017-09-10T16:06 flare event suggests that beam 
heating is continuous for at least an hour after the flare onset. 
This produces a sequential heating of the flare loop assembly, observed 
as an apparent expansion of the loop arcade. RADYN cannot provide 
such a multithread approach as it simulates the flare as a single loop event. 
Therefore, we simulate the flare profiles observed at a footpoint of a specific 
loop at a specific time, with the RADYN model incorporating an instant 
electron beam injection lasting for 20 sec along the single loop.

\subsection{Synthetic \hb\ and \cawav\ Line Profiles}

RADYN simulations yield synthetic \hb\ and \ca\ line profiles
for the flare model F11 at five different directions from
disk center at $\mu=0.95$ to the extreme limb at $\mu=0.047$ shown
in Figure~\ref{fig5}. 
This particular set of $\mu$ is used in RADYN for the angular
integration of the radiation field, which is anisotropic due 
to velocities. Due to the large optical path length, the optical depth is 
much larger towards the limb than for the vertical direction. 
As a result \hb\ develops a strong central reversal encompassed 
by two emission peaks, whose intensity decreases toward the limb 
(Figure~\ref{fig5}: left panel). Conversely, the \ca\ line core 
intensity increases toward the limb, however, at $\mu=0.047$ 
it decreases and develops a weak central 
dip (Figure~\ref{fig5}: right panel) indicating that 
the optical depth is not changing significantly with respect to $\mu$ 
for the emission near the line core, whereas the source function is 
increasing dramatically (see a detailed explanation in section~\ref{temop})}.
The simulations show that the line wings have 
lower opacity and higher emissivity towards the limb for both lines 
(Figure~\ref{fig5}).

\begin{figure}
\includegraphics[width=0.349\textwidth]{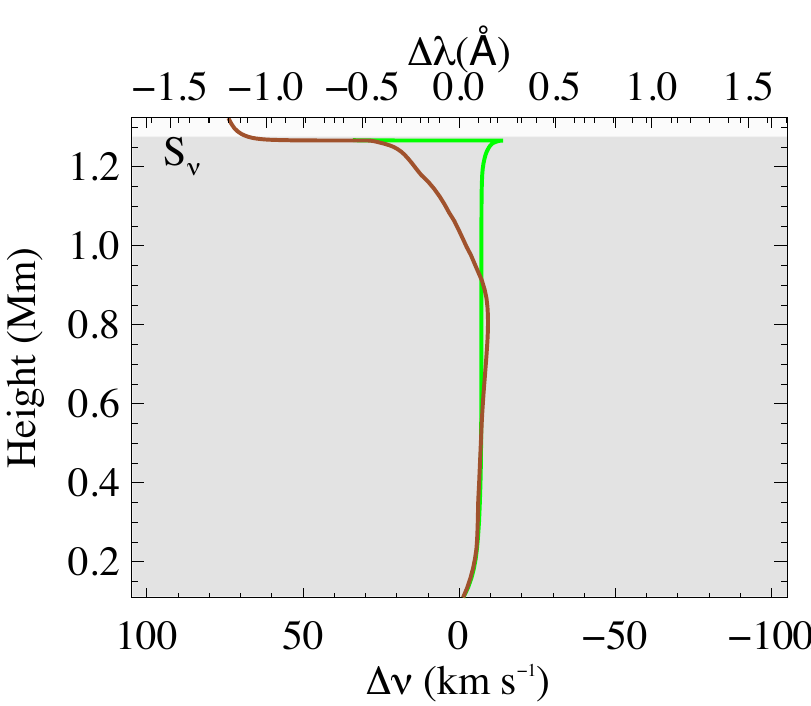}
\includegraphics[width=0.323\textwidth]{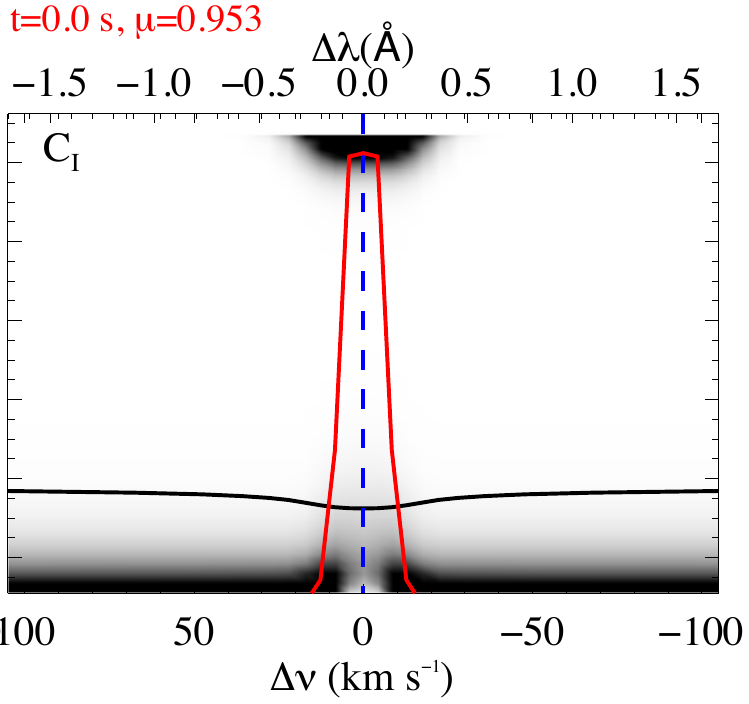}
\includegraphics[width=0.323\textwidth]{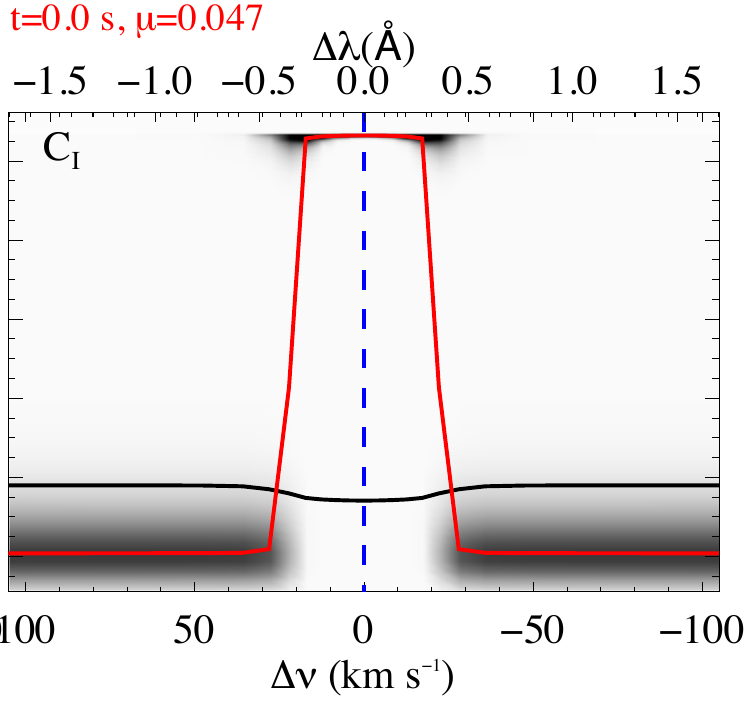}
\includegraphics[width=0.349\textwidth]{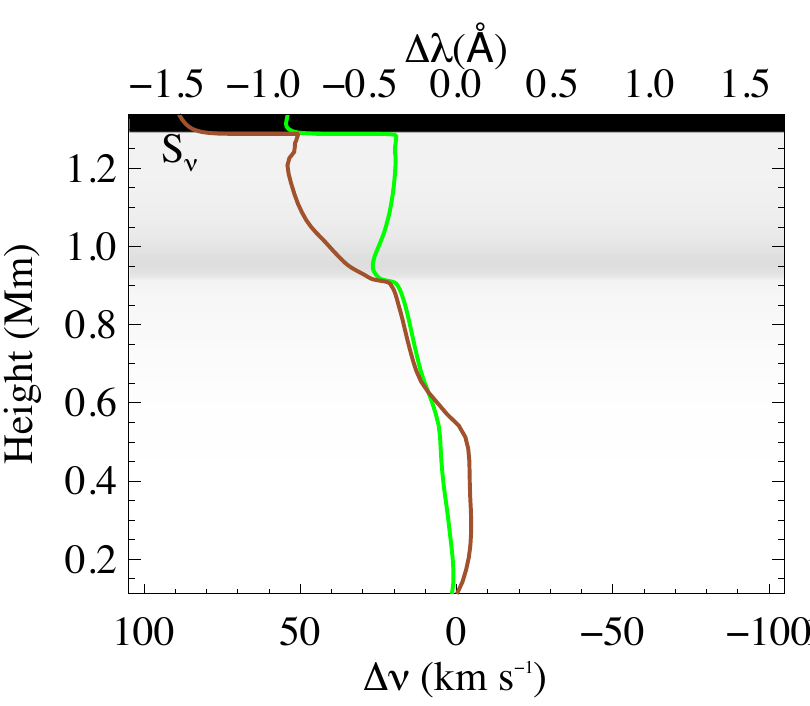}
\includegraphics[width=0.323\textwidth]{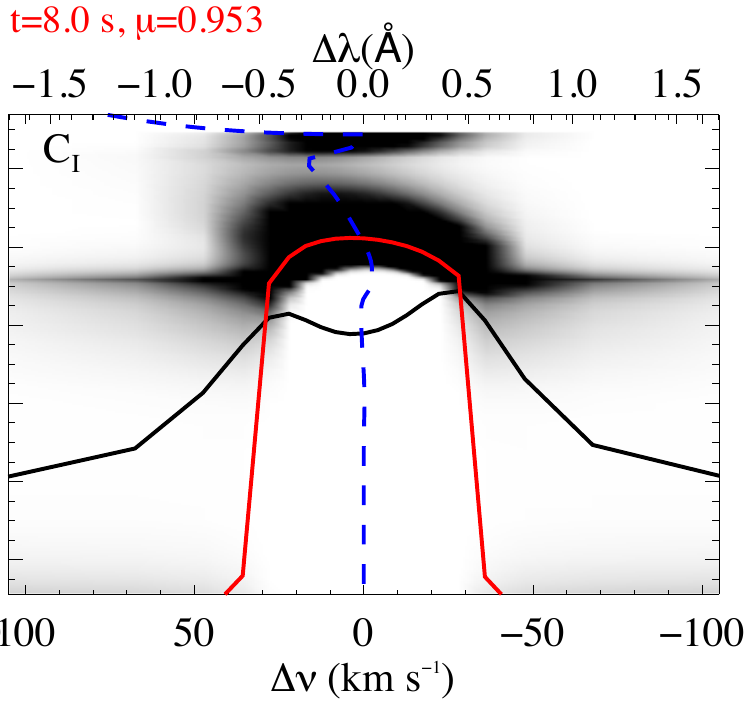}
\includegraphics[width=0.323\textwidth]{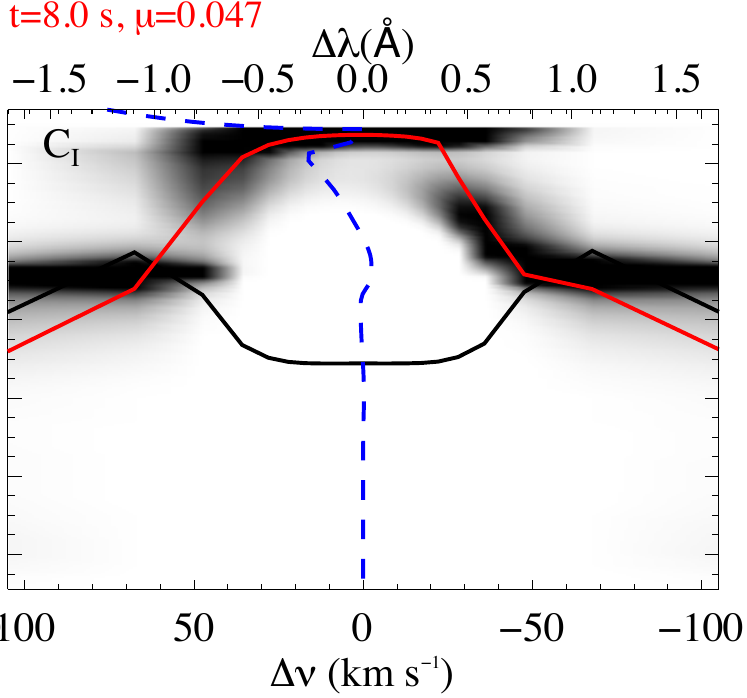}
\caption{Source function and intensity contribution functions $C_I$ of the
  \hb\ line for the direction of $\mu$=0.047 and $0.953$. Top three panels:
  pre-flare $C_I$ at $t=0$\,s. Bottom 3 panels: 8\,s after onset of
  the F11 flare. Darker shades indicate higher values. The \hb\ line
  profile is overplotted in the $C_I$ panels as a black line in an arbitrary scale. Red lines indicate the  $\tau =
  1$ layer. The vertical velocity component $v_{\rm z}$ is overplotted as
  a blue dashed line where a positive value indicates an upflow. The line
  source function S$_\nu$ (green line) and the Planck function (brown
  line) are overplotted in the left panels.}
\label{fig6}
\end{figure}

\begin{figure}
\includegraphics[width=0.349\textwidth]{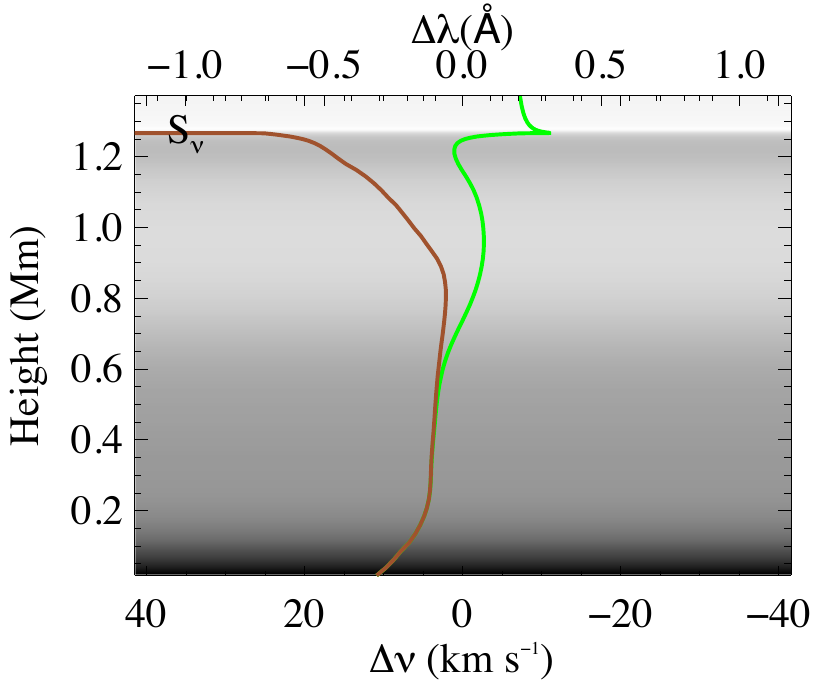}
\includegraphics[width=0.323\textwidth]{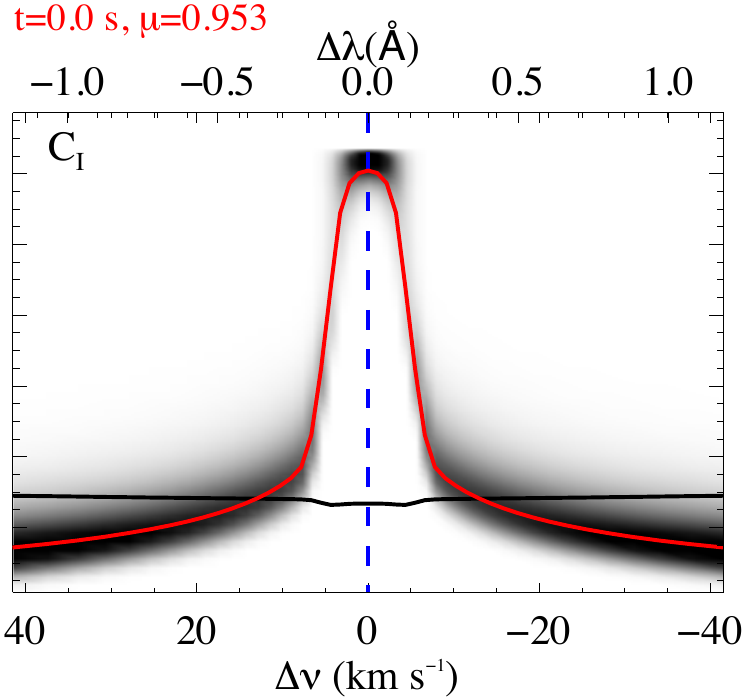}
\includegraphics[width=0.323\textwidth]{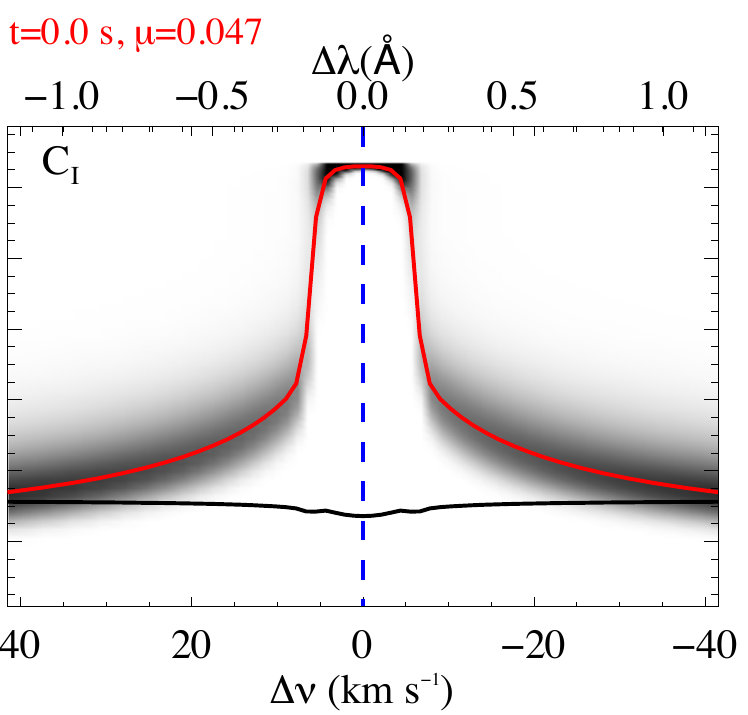}
\includegraphics[width=0.349\textwidth]{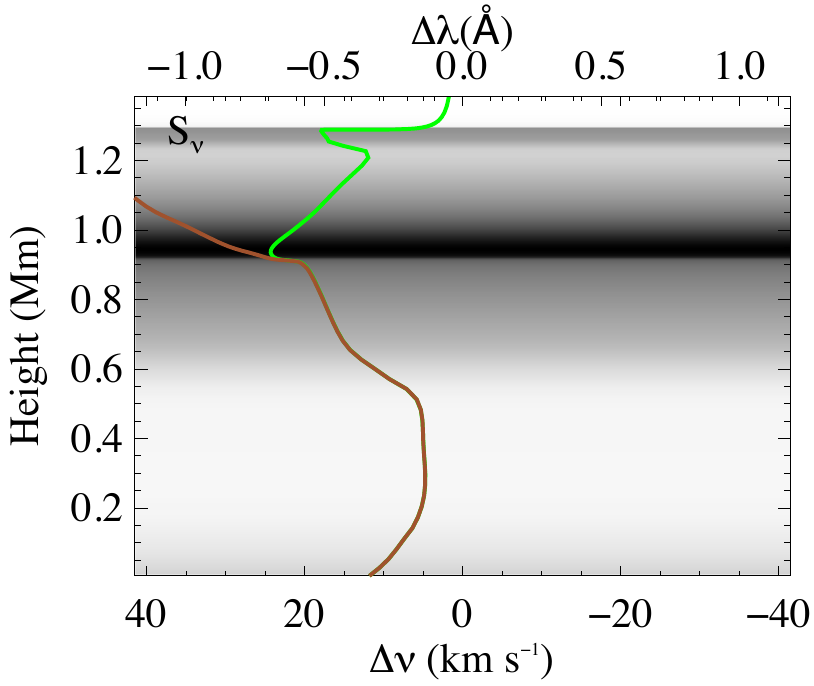}
\includegraphics[width=0.323\textwidth]{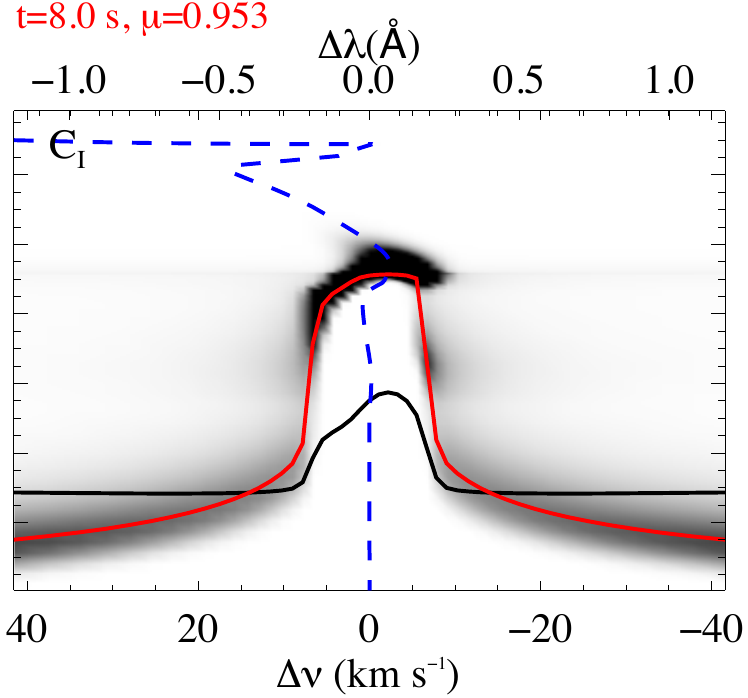}
\includegraphics[width=0.323\textwidth]{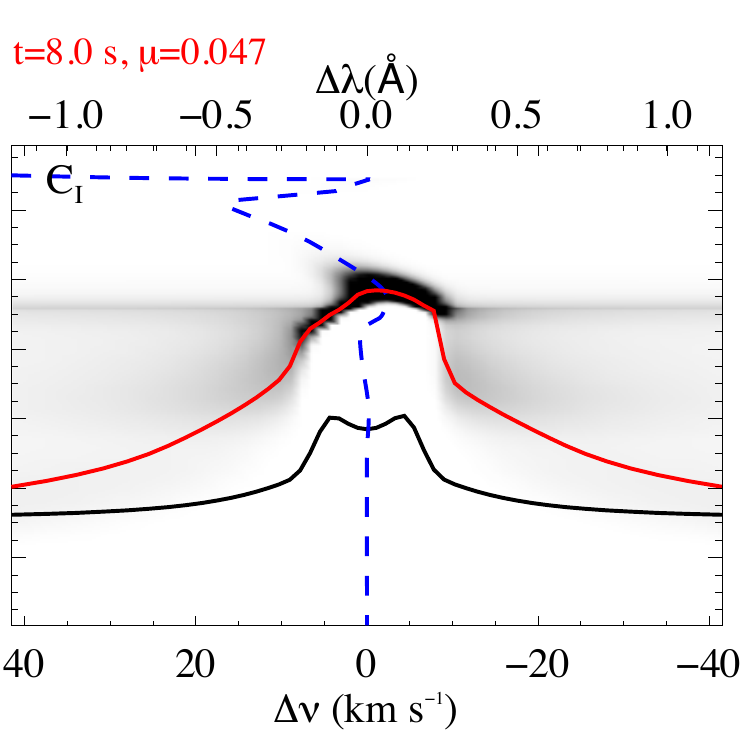}
\caption{Same as Figure~\ref{fig6}, but for \cawav.}
\label{fig7}
\end{figure}

\subsection{Contribution Function and Formation Heights}

To determine the formation heights of the \hb\ and \ca\ lines in the 
flaring atmosphere, we examine their intensity contribution functions
$C_I$, which show the intensity as a function of wavelength and as a
function of distance along the LoS. \citet{CarlssonStein1997} introduced a formal
solution of the radiative transfer equation for the emergent intensity
expressed in terms of component factors of $C_I$ in the form:
\begin{equation}
\label{eq1}
I_\nu = \int_{z}C_I\,dz/\mu = \int_{z}S_\nu\,\chi_\nu\exp(-\tau_\nu/\mu)\,dz/\mu\,=\int_{\tau_\nu} S_\nu\,\exp(-\tau_\nu/\mu)\,d\tau/\mu,
\label{eq1}
\end{equation}
where $z$ is the atmospheric height normal to the photosphere 
(z=0 is at $\mathrm{\tau_{5000}}$=1), 
and $S_\nu$, $\tau_\nu$, and $\chi_\nu$ are the source function, 
optical depth, and opacity (linear extinction coefficient) 
along the vertical direction, respectively. 

In Figures~\ref{fig6} and \ref{fig7} we present $C_I$ and $S$ diagrams 
for the \hb\ and \ca\ lines. The depicted quantities are plotted in inverse grayscale 
with darker areas showing higher values. Figure~\ref{fig6} shows the $C_I$
of the pre-flare \hb\ line profile at $t =0$\,s (top panels) and 
8\,s (bottom panels) after the onset of beam heating by
the F11 flare model near the limb and near the disk center at the direction
$\mu = 0.047, 0.953$, respectively. The vertical axes in 
Figures~\ref{fig6} and \ref{fig7} show $z$, the height normal to the 
photosphere ($\mu=1$). Therefore, the contribution function 
computed along the LoS is given as a function of the 
projected LoS distance with respect to the vertical height in 
Figures~\ref{fig6} and \ref{fig7}. In RADYN the $z \sim$\,0\,Mm is 
defined as the base of the photosphere at $\tau_{\rm 5000} = 1$. 
The bottom middle and right panels of Figure~\ref{fig6} show that during 
the beam heating the $\tau = 1$ layer at \hb\ line wings is formed much
higher than in the pre-flare atmosphere (the top middle and right 
panels of Figure~\ref{fig6}). Consequently, the near-limb \hb\ line wing 
intensities are formed much higher in the chromosphere at the heights of 
$\sim 0.9-0.95$\,Mm compared to the pre-flare
intensity, which is formed at $\sim 0.1-0.35$\,Mm. The maximum
of $C_I$ of the \hb\ wing intensities occurs above the $ \tau = 1$
layer during the beam heating (Figure~\ref{fig6}:
bottom right $C_I$ panel). Furthermore, due to the large optical path 
length, the optical depth is much larger along the direction of lower $\mu$. 
Therefore, the $\tau = 1$ layer is formed much higher for the near-limb 
flare emission than near the disk center (Figure~\ref{fig6}). 
This suggests that, at $\mu$=0.047, the emergent intensities 
reflect conditions in a higher layer above $\mathrm{\sim800~km}$, 
contrary to the case of $\mu$=0.95 where the line profile spans a
wide range of the lower flare atmosphere. 

The difference between the $\tau = 1$ layers in the pre-flare and flare  
atmosphere for the different viewing geometries also changes the
formation heights for the \ca\ emissions 
(Figure~\ref{fig7}). The $C_I$ diagrams at $t =0$ and 8\,s show
that the $\tau = 1$ layer for \ca\ line wings are formed at similar 
heights for the same $\mu$ in the pre-flare and flare atmosphere (Figure~\ref{fig7}). 
However, similar to \hb\ line, the $\tau = 1$ layer for the \ca\ line wings 
is formed higher for near limb emission ($\mu=0.047$) compared  
to the $\tau = 1$ at the near disk emissions ($\mu=0.953$) (Figure~\ref{fig7}).

The left panels in Figures~\ref{fig6} and \ref{fig7} show that
the line source functions of \hb\ and \ca\ are affected by the
electron beam heating. For \ca, the beam heating moves the 
point where the source functions decouple from the Planck function 
from heights of 0.6\,Mm to 0.95\,Mm. However, because the decoupling 
occurs above the formation heights of the \ca\ line wings these are still 
formed under LTE conditions.

\begin{figure}
\includegraphics[width=\textwidth]{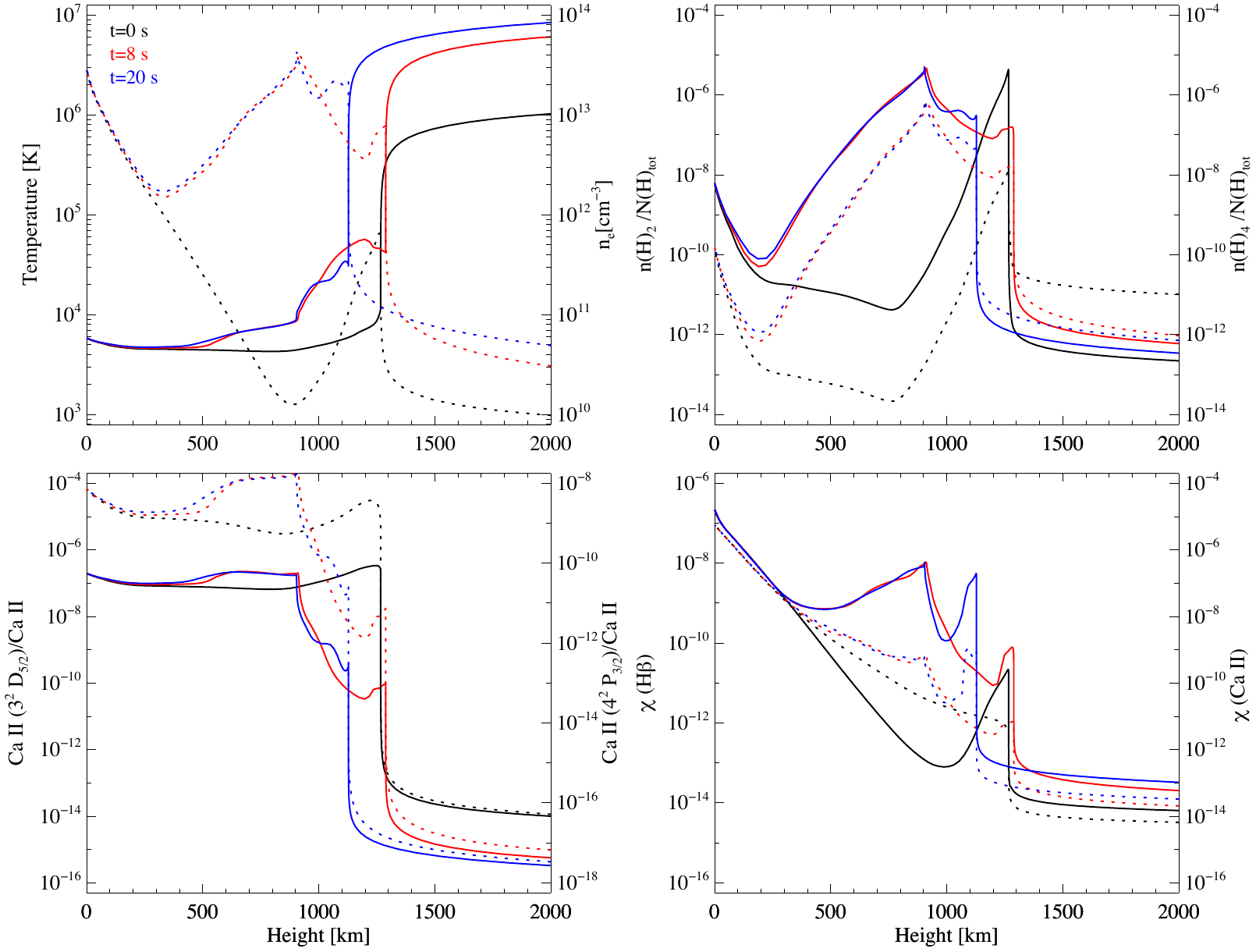}
\caption{The temporal evolution and height distribution of the atmospheric
  parameters (top left) and line formation characteristics of the
  \hb\ and \cawav\ lines. Pre-flare values at
  $t=0$\,s are shown in black. The corresponding profiles 
  at 8\,s and 20\,s after the onset
  of the F11 flare are shown in red and blue, respectively. The solid
  and dotted lines correspond to the left and right $y$ axis,
  respectively. Top left: Temperature (left $y$ axis) and electron
  density (right $y$ axis). Top right: The relative population
  densities of the $n=2$ (left $y$ axis) and $n=4$ (right $y$ axis)
  levels of hydrogen with respect to the total number density of
  hydrogen (neutrals). Bottom left: The relative population densities 
  of the $3\,^2{\rm D}_{5/2}$ (left $y$ axis) and $4\,^2{\rm P}_{3/2}$ 
  (right $y$ axis) energy states with respect to the density
  of the \ion{Ca}{2} ions (right $y$ axis). Bottom right: Line wing
  opacities $\chi$ of \hb\ at $\Delta\lambda= -1$\,\AA\ (left $y$
  axis) and \cawav\ at $\Delta\lambda= -0.6$\,\AA\ (right $y$ axis)
  from the line center.}
\label{fig8}
\end{figure}

\subsection{Temperature, Electron and Population Densities and Opacities}
\label{temop}
Figure~\ref{fig8} shows the temporal evolution and height distribution of
the atmospheric parameters and line formation characteristics of the
\hb\ and \ca\ lines covering the pre-flare stage at $t = 0$\,s and at
two heating stages of $t = 8$\,s and 20\,s after the onset of the electron
beam heating. During the beam heating the temperature
increases to $\sim 5000-30000$\,K over
heights of $\sim 300-1100$\,km (Figure~\ref{fig8}: top left panel), 
accompanied with an increase in collisional rates and electron density
$n_{\rm e}$, from about $5 \times 10^{10}$\,\cmcub\ up to $\sim 4 \times
10^{13}$\,\cmcub.


The electron beam and increased temperature leads to a rapid change in
the populations of atomic levels. The \hb\ line
arises from transitions between the $n = 2$ and $n = 4$ levels of
neutral hydrogen. The beam heating excites electrons from the ground state 
and the first excited states to the upper levels: the collisional excitation 
(ionization) is due to thermal collisions (enhanced $T$ and $n_e$ due to
heating) and direct non-thermal excitation (ionization) by the beam electrons.
The top right panel of Figure~\ref{fig8} shows that the ratio of 
the $n = 2$ level population over the total number of hydrogen atoms increases
significantly during the beam heating over the height range
$300-1100$\,km, invoking an increase of opacity in the \hb\ line wing
(Figure~\ref{fig8}: bottom right panel). Consequently, 
the probability that the \hb\ wing photons from deep layers at $\sim
100-400$\,km\ will be
absorbed by the overlying atmosphere is increased. Therefore the \hb\ wing
intensity is formed in the higher layers of the flaring atmosphere
compared to the pre-flare stage (Figure~\ref{fig6}). The population of the
$n = 4$ hydrogen level is also enhanced due to the beam heating
(Figure~\ref{fig8}: top right panel) invoking an increase of the
source function and line intensities. We note that the line source function 
is frequency independent due to the assumption of CRD.
 
The ratio of the total population densities \ion{Ca}{2}/\ion{Ca}{3}
(not shown) of two successive ionization stages of calcium decreases
significantly at heights below $\sim$\,1250\,km after the onset of
the electron beam heating, suggesting an increased density of calcium
ions in the second ionization stage.  The \ca\ line arises due to 
transitions between the $3\,^2{\rm D}_{5/2}$ and $4\,^2{\rm P}_{3/2}$ 
energy states.  The bottom left panel of
Figure~\ref{fig8} shows that despite the dramatic decrease of population
density of the \ca\ ions, the population density of the $3\,^2{\rm D}_{5/2}$ 
state does not change much over the height range
$\sim200-900$\,km, which corresponds to the formation heights of the
\ca\ line wings (Figure~\ref{fig7}: bottom right $C_I$
panels). Therefore the \ca\ line wing opacity changes only slightly
over these heights (Figure~\ref{fig8}: bottom right panel). On the
contrary, the heating invokes an increase of two orders of magnitude 
in the \hb\ line wing opacity over the heights $\sim 500-1000$\,km
(Figure~\ref{fig8}: bottom right panel). However, the population
density of the \ca\ $4\,^2{\rm P}_{3/2}$ state increases by a factor of 10 over
the height range $\sim200-900$\,km, invoking an increase of the
source function and line intensities at the beam heating phase 
(Figures~\ref{fig7}: bottom right $C_I$ panel).

\begin{figure}
\includegraphics[width=0.495\textwidth]{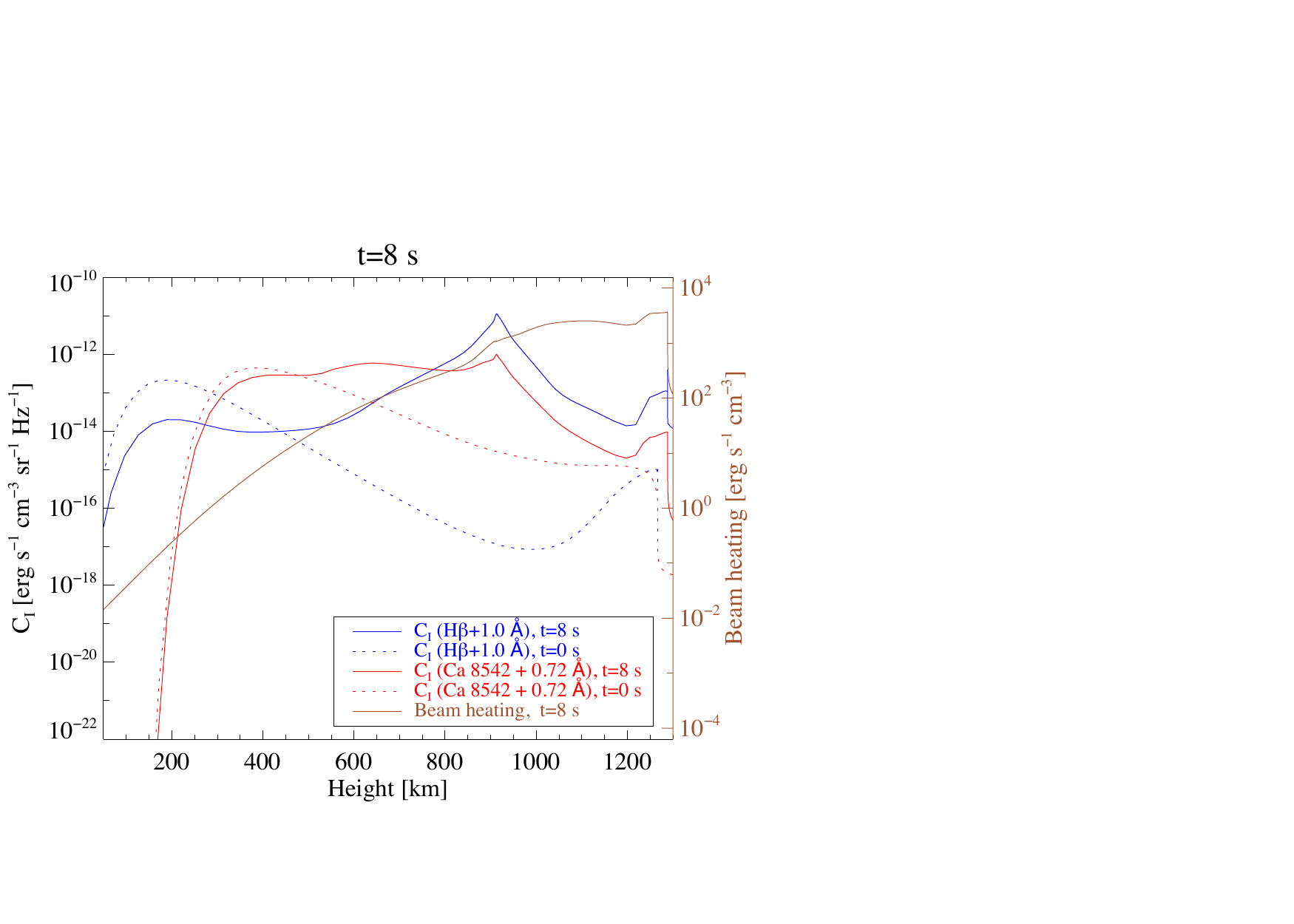}
\includegraphics[width=0.495\textwidth]{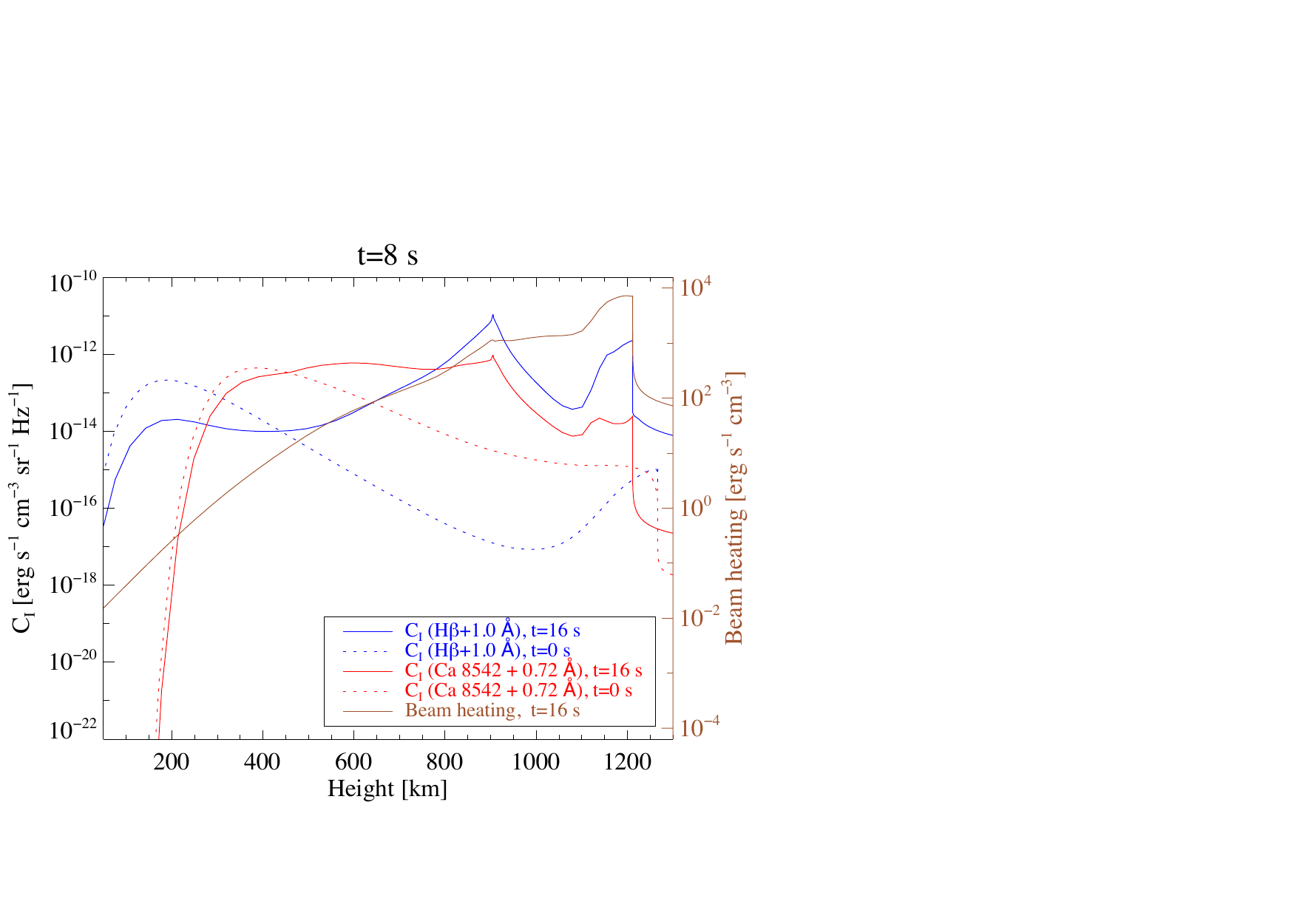}
\caption{Temporal evolution and height distributions of beam heating
  (brown lines) and the intensity contributions function $C_I$ along the  
  LoS tin the \hb\ line wing at $\Delta\lambda = +1.0$\,\AA\ (blue lines)
  and in the \cawav\ line wing at $\Delta\lambda = +0.72$\,\AA\ (red
  lines) from the line center. Pre-flare $C_I$ at $t=0$\,s are
  shown as dotted lines in both panels. Beam heating and $C_I$
  8\,s and 16\,s after the onset of the F11 flare are
  shown in the left and right panel, respectively.}
\label{fig9}
\end{figure}

\subsection{Beam Heating versus Contribution Functions}

Figure~\ref{fig9} shows the beam energy deposition together with 
$C_I$ (along the LoS) of the \hb\ and \ca\ line wing intensities at 
$\Delta\lambda =1.0$\,\AA\ and $0.72$\,{\AA}, respectively as a function 
of height for the pre-flare atmosphere at $t = 0$\,s and during the heating 
at $t =8$\,s (left panel) and $t = 16$\,s (right panel). The panels show 
that the \hb\ $C_I$ at $\Delta\lambda = 1.0$\,\AA\ has a pre-flare
maximum at $\sim$\,150\,km. However, during the beam heating the
maximum raises to $\sim$\,900\,km and the \hb\ $C_I$ curve declines
exponentially towards the lower layers.  
The contribution functions, computed from 
Equation~\ref{eq1}, are defined by the opacity and emissivity, and 
hence the population densities of the atmosphere along the LoS. 
As shown in Figure~\ref{fig8}, the population densities of 
the hydrogen $n = 2$ and $n = 4$ levels and line wing opacities 
decrease exponentially during the beam heating over the height range
$900-200$\,km (Figure~\ref{fig8}: right panels).
The beam heating function also decreases 
exponentially along the lower solar atmosphere (Figure~\ref{fig8}). 
As a result, the decline in contribution function
follows closely the declining energy deposition function over heights
$\sim 900 - 500$\,km, suggesting that the \hb\ line wing intensities
might be a proxy for beam energy deposition in the flaring atmosphere (Figure~\ref{fig8}).
Along $\sim 900 - 500$\,km the Planck function and 
source function are still not decoupled, indicating that the H$\beta$ wing is 
forming under LTE conditions (Figure~\ref{fig6}). Similar to the contribution function, the 
population densities, opacities, beam heating function,  and
the temperature also decrease exponentially over this height range (Figure~\ref{fig8}: top left panel). 
This suggests that the beam energy over these heights is consumed by the heating,
and heating (temperature) is the main parameter that redistributes the 
population densities/opacities and defines the formation of the 
H$\beta$ line (emergent intensities).

The $C_I$ of the \ca\ wing intensity at
$\Delta\lambda = 0.72$\,\AA\ shows that there is no obvious 
correlation with the beam energy deposition curve.
This is because the electron beam heating ionizes \ion{Ca}{2} to \ion{Ca}{3} 
rather then populating excited energy states responsible for 
\ion{Ca}{2} line emission (Section~\ref{temop}).

\section{Discussion and Conclusions}

Observations of an X8.2-class solar flare,
acquired with high spatial resolution in the \hb\ and \ca\ lines, are
compared with the results of RHD simulation
with the model parameters based on an analysis of HXR data. 
The observations give a rare off-limb viewpoint of flare ribbons, 
which are the flare-associated emissions at the footpoints of 
flare loops in the line wing images of \hb\ and \ca\ lines
(Figures~\ref{fig1}--\ref{fig3}). The line profiles of the flare ribbons 
are broad, with significantly increased emissions in the line 
wings (Figures~\ref{fig2}--\ref{fig3}: bottom panels). 
Far wing flare intensities of \hb\ and \ca\ lines can be considered as 
emission from an optically thin plasma, hence they are less 
attenuated by the overlying atmosphere along the LoS
\citep{Asaietal2012}. On the other hand, we know that the horizontal
dimension of the flare ribbons is extended across far larger 
distances than the vertical width. Therefore, at the limb, there 
is far more optically thin ribbon plasma emitting along the LoS 
compared to a disk center viewing geometry. This produces an 
observed enhancement of the line wing intensities for off-limb 
flare ribbons. The analysis of synthetic flare line profiles from 
the RADYN model at different heliocentric angles confirms the 
enhancement of the far wing emissions of \hb\ and \ca\ lines 
toward the limb (Figure~\ref{fig5}). One of the main 
limitations of the RADYN model used in the present work 
is that it considers the flare as a single
loop, short-duration event whereas the observed flare 
arcade consists of many independent loops heated sequentially 
with continuous reconnection processes. Multithread hydrodynamic flare 
simulations presented in \cite{Warren2006} have shown that the 
multithread modelling approach is an important consideration 
that can resolve some basic inconsistencies reported between 
observations and models (e.g. slow decay of the observed flare emission).  
The results from \cite{Warren2006} suggest that the heating timescale 
for individual loops could be on the order of 200 s, 
suggesting that in multithread flares, heating could last tens of 
minutes or even an hour. With RADYN we simulate the flare profiles 
observed at the footpoint of a specific loop at a specific time 
with the model which incorporates an electron beam injection 
pulse lasting for 20 sec along the loop.
Alternatively, the observed flare ribbon and extended 
HXR emission can be investigated using the models representing 
the atmosphere at the cooling phase preceded by intense electron 
beam heating. It was shown by F-P simulations that the beam 
electrons can become trapped at the stopping depth of lower 
energy electrons for an extended period of time \citep{Siverski2009}. 
The appearance of the observed off-limb flare ribbon can be 
associated to these depths.

The synthetic line profiles in Figure~\ref{fig5} reveal that the
intensity near $\Delta\lambda = \pm 1.2$\,\AA\ at $\mu = 0.047$ is
about four times higher than the disk center continuum near \hb,
whereas the intensity in \ca\ line wings is 
only about 1.5 times higher than the disk center continuum near \ca. 
The \hb\ and \ca\ line wing ribbons have
different contrast, and hence different visibility, near the
limb. Indeed, our off-limb observations show that the ribbons are
far brighter and more prominent in the \hb\ wing images than in the \ca\ wing
images. \citet{Avrettetal1986} also showed similar differences
between synthetic \ha\ and \ion{Ca}{2}~8498\,\AA\ line profiles at
different $\mu$ (see pages 238 and 243 therein).
The main reason for this difference is that due to the low 
ionization energy, the fraction \ion{Ca}{2} to \ion{Ca}{3} during 
the beam heating decreases by approximately 10 to 100 
times in the lower solar atmosphere indicating that a significant 
number of  \ion{Ca}{2} ions are becoming ionized to \ion{Ca}{3}. 
Therefore, in contrast to Balmer lines, the beam precipitation ionizes 
\ion{Ca}{2} to \ion{Ca}{3} rather than producing excited levels of \ion{Ca}{2}.

Ribbons are absent in the \hb\ and \ca\ line core and near-core images
(Figure~\ref{fig1}: left panels). Their absence is due to the maximum 
opacity at the line cores: at $\mu\sim$\,0, the line core radiation can
not transmit along the LoS through the opaque chromosphere canopy. 
Unfortunately, this is not captured in the RADYN model used in
this work. The synthetic line profiles show enhanced line core
intensities with strong central reversal in \hb\ and at most a very
weak central reversal in \ca. RADYN uses a 1D plane-parallel
geometry and therefore, at small values of the direction
cosine $\mu$, the emergent intensity from the model
is calculated by integration of the source function over large distances 
through the flare atmosphere, compared to the curved plane of the true 
chromosphere where the flare only exists in a localised area.  
In addition, the observed LoS is populated by optically thick 
chromospheric structures (e.g., spicules, prominences) 
which are, of course, absent from the model.  
Therefore, the model overestimates the ribbon emissions, and the line core in
particular.

The absence of ribbons in the \ca\ far-wing images at
$\Delta\lambda = \pm 1.75$\,\AA\ (Figure~\ref{fig3}: top right
panel) can be explained by a simple geometrical effect. 
The photospheric footpoints of the observed flare
loops are not detected in the SST dataset since they are behind the
limb \citep{Kuridzeetal2019}.  Therefore, the ribbons are also located
behind the limb. However, due to the elevation we can see them, 
at least partially, off-limb. In the \hb\ line wing, the apparent separation, 
or the projected formation height of the ribbon in the plane of the sky with 
respect to the nominal limb, is about $300-500$\,km. 
The \ca\ ribbon at $\Delta\lambda = \pm 0.8$\,\AA\ does not 
show any clear separation from the \ca\ limb
defined as the edge of the solar disk at $\Delta\lambda = \pm
1.75$\,\AA. The ribbons at the line wing positions outside
$\Delta\lambda = \pm 0.8$\,\AA\ are located deeper in
the atmosphere and are obscured by the limb. 

To understand the formation of the flare ribbons, we analyze
components of the intensity contribution functions for the synthetic
\hb\ and \ca\ lines computed with the RADYN code. We find that the 
near-limb maximum \hb\ line wing emissions, interpreted 
as flare ribbons, are formed in the atmosphere at around 
950\,km above the base of the photosphere 
at $\tau_{\rm 5000} = 1$ (Figure~\ref{fig6}: bottom 
right $C_I$ panel).  In the non-flaring atmosphere \hb\ line wing 
emissions are formed much deeper, at a height of
around 200\,km (Figure~\ref{fig6}: top right $C_I$
panel). The beam heating rapidly changes the balance between the
hydrogen populations in the
chromosphere, with increased collision rates 
(both thermal and non-thermal) exciting electrons from
the lower to the upper levels (Figure~\ref{fig8}: top right panel). 
This leads to a significant increase in the \hb\ line wing opacities 
between heights of $\sim 400-950$\,km above the photospheric limb
(Figure~\ref{fig8}: bottom right panel).  As a result, \hb\ line wing
photons emitted at a height of around 200\,km are no
longer able to escape freely along the LoS, which results in a 
change of formation height.

Our analysis indicates that the balance between the population densities 
of the energy states responsible for the \ca\ emission is also affected 
by the rapid flare heating (Figure~\ref{fig8}: bottom left panel).  
Opacities of the \ca\ line wings change only slightly by the beam heating
(Figure~\ref{fig8}: bottom right panel). However, due to the
significant increase in the upper-level population density of the 
\ca\ line, the emissivity and the source function are increased in the
upper layers.  As a result the formation heights of the \ca\ line wings
are extended up to $\sim$\,900\,km during the beam heating phase
(Figure~\ref{fig7}: lower right $C_I$ panel).

The observed separation between the off-limb ribbons and
the limb photosphere is well reproduced by the simulations. The
observed apparent separation (i.e., projection of the formation
heights in the plane of sky) between the ribbon and the nominal limb
is $\sim 300-500$\,km for \hb\ and $\sim$\,0\,km for \ca. They
are smaller than the $C_I$-based formation heights in the model
($\sim$\,900\,km for \hb\ and $\sim$\,400\,km for \ca) measured with
respect to the base of the photosphere at $\tau_{\rm 5000} = 1$. The
difference is likely due to uplift of the nominal limb above the
base of the photosphere by about 350\,km
\citep[Table~1]{Lites1983}. Subtracting this offset from the
$C_I$-based heights brings them to a good agreement with the
observed ribbon-limb separations.
The comparison of the beam energy deposition and the \hb\ intensity
contribution function at $\pm 1.0$\,\AA\ shows that in the current model 
the \hb\ line wing intensities are a good proxy for the beam penetration
and flare energy deposition over heights in the lower solar atmosphere
(Figure~\ref{fig9}). 

To our knowledge, this is the first imaging spectroscopy at
high spatial resolution of off-limb solar flare ribbons in the
hydrogen \hb\,4861\,\AA\ Balmer line and the near-infrared line of
single ionized calcium \cawav. Emission in the
optically thin wings of the \hb\ line show high-contrast
off-limb flare ribbons which are elevated significantly above the photosphere.
The results are very encouraging for future off-limb
flare studies using chromospheric diagnostics, including 
the advanced modelling encompassing the opaque chromospheric
canopy and spicules. The new-generation 4-m Daniel
K.~Inouye Solar Telescope (DKIST), with its advanced
spectropolarimetric facilities, will offer excellent opportunities
for off-limb flare measurements which, as demonstrated here, can
provide powerful diagnostics of the flare atmosphere.

\acknowledgments

The research leading to these results has received funding from the
S\^{e}r Cymru II scheme, part-funded by the European Regional Development Fund
through the Welsh Government, and STFC grant ST/S000518/1 to Aberystwyth University.
The work of D.K. was supported by Georgian Shota Rustaveli 
National Science Foundation project FR17\_\ 323.
MM acknowledges support from STFC through grant number:
ST/P000304/1. The Swedish 1-m Solar Telescope is
operated on the island of La Palma by the Institute for Solar Physics
of Stockholm University in the Spanish Observatorio del Roque de los
Muchachos of the Instituto de Astrof\'{\i}sica de Canarias. The
Institute for Solar Physics is supported by a grant for research
infrastructures of national importance from the Swedish Research
Council (registration number 2017-00625). 
P.H. acknowledges support from the Czech Funding Agency grant 19-09489S.
J.K. acknowledges the project VEGA 2/0048/20. R.O. acknowledges support from the Spanish 
Ministry of Economy and Competitiveness (MINECO) and FEDER funds 
through project AYA2017-85465-P. We would like to 
thank the anonymous referee for comments and 
suggestions that helped improve this manuscript.

\facility{SST(CRISP,CHROMIS)}

\bibliography{dkuridze_etal_revised}{}
\bibliographystyle{aasjournal}

\end{document}